\documentclass[a4paper,11pt]{article}
\usepackage{jcappub}
\usepackage[utf8]{inputenc}
\usepackage{amsmath}
\usepackage{amssymb}
\usepackage{graphicx}
\usepackage{xcolor}
\usepackage{ulem}
\usepackage{booktabs}
\usepackage{algorithm}
\usepackage{algpseudocode}
\usepackage{hyperref}
\usepackage{aas_macros}
\usepackage{subcaption}
\usepackage{float}
\usepackage[T1]{fontenc}
\usepackage[frozencache=true,cachedir=minted-cache]{minted}
\usepackage[british]{babel}

\newcommand{\LCDM}{$\Lambda$CDM}

\newcommand*{\TTTEEE}{\ensuremath{T\!T/T\!E/E\!E}}

\newcommand*{\planck}{\textit{Planck}}

\newcommand*{\ole}{\texttt{OLÉ}}
\newcommand*{\candl}{\texttt{candl}}
\newcommand*{\montepython}{\texttt{MontePython}}
\newcommand*{\cobaya}{\texttt{Cobaya}}
\newcommand*{\class}{\texttt{CLASS}}
\newcommand*{\camb}{\texttt{CAMB}}
\newcommand*{\getdist}{\texttt{GetDist}}
\newcommand*{\arviz}{\texttt{ArviZ}}

\newcommand{\algorithmictype}[1]{\texttt{#1}}

\newcommand{\oleparam}[2]{\texttt{#1 :} \textbf{#2}}

\title{\boldmath OLÉ - Online Learning Emulation in Cosmology}

\author[a]{Sven Günther,}
\emailAdd{sven.guenther@rwth-aachen.de}

\author[b]{Lennart Balkenhol,}
\emailAdd{lennart.balkenhol@iap.fr}

\author[a]{Christian Fidler,}
\emailAdd{fidler@physik.rwth-aachen.de}

\author[b]{Ali Rida Khalife,}
\emailAdd{ridakhal@iap.fr}

\author[a]{Julien Lesgourgues,}
\emailAdd{lesgourg@physik.rwth-aachen.de}

\author[a]{Markus R. Mosbech,}
\emailAdd{mosbech@physik.rwth-aachen.de}

\author[a]{Ravi Kumar Sharma}
\emailAdd{rksharma@physik.rwth-aachen.de}

\affiliation[a]{Institute for Theoretical Particle Physics and Cosmology  (TTK),
RWTH Aachen University, Sommerfeldstr. 16, D-52056 Aachen, Germany}
\affiliation[b]{Sorbonne Universit\'{e}, CNRS, UMR 7095, Institut d'Astrophysique de Paris, 98 bis bd Arago, 75014 Paris, France}

\abstract{In this work, we present \ole{}, a new online learning emulator for use in cosmological inference. The emulator relies on Gaussian Processes and Principal Component Analysis for efficient data compression and fast evaluation. Moreover, \ole{} features an automatic error estimation for optimal active sampling and online learning.
All training data is computed on-the-fly, making the emulator applicable to any cosmological model or dataset.
We illustrate the emulator's performance on an array of cosmological models and data sets, showing significant improvements in efficiency over similar emulators without degrading accuracy compared to standard theory codes.
We find that \ole{} is able to considerably speed up the inference process, increasing the efficiency by a factor of $30-350$, including data acquisition and training. Typically the runtime of the likelihood code becomes the computational bottleneck.
Furthermore, \ole{} emulators are differentiable; we demonstrate that, together with the differentiable likelihoods available in the \candl{} library, we can construct a gradient-based sampling method which yields an additional improvement factor of 4.
\ole{} can be easily interfaced with the popular samplers \montepython{} and \cobaya{}, and the Einstein-Boltzmann solvers \class{} and \camb{}.
\ole{} is publicly available at \url{https://github.com/svenguenther/OLE}.
}

\begin{document}
\maketitle
\flushbottom

\section{Introduction}
\label{sec:intro}
With the ever increasing precision and complexity of data sets gathered by modern cosmological experiments \cite{LSSTScience:2009jmu,ACT:2020frw,DESI:2024mwx,Planck:2019nip,SPT-3G:2022hvq,EuclidTheoryWorkingGroup:2012gxx,SKA:2018ckk}, the requirements for computational accuracy rise in tandem. For the use of these data sets to remain feasible for inference, we must also improve the efficiency with which we compute our theoretical predictions without degrading their accuracy. Emulators are a common solution to this problem, allowing for fast calls of a function trained to reproduce the output of a more computationally expensive theory code. These are often pre-trained {\cite{cosmopower22,Euclid:2018mlb,Arico:2020lhq,Arico:2021izc,Moran:2022iwe}, relying on a library of pre-computed calls to the theory code for a variety of input parameters. This offers the advantage of fast predictions out of the box, at the cost of flexibility, as the emulator can only be assumed to be accurate within the parameter space it is trained on. In particular, this is a problem in cases where the relevant parameter space or the required accuracy is a priori unknown, leading to overly conservative coverage of the parameter space. This also limits the usefulness of pre-trained emulators when considering models outside of the standard Cosmological Constant + Cold Dark Matter ($\Lambda$CDM) cosmology and its most common extensions.

In this work, we present an alternative type of emulator, relying on online learning\footnote{Online learning is a training strategy relying on training data becoming available sequentially, allowing the model to be trained progressively as more data becomes available. One example is by training on (a subset of) steps in a sampling algorithm (see~\cite{ML_Foundations,Gunther:2023xhh} for more details).} to work jointly with the theory code to produce training data and train on-the-fly for whichever model and parameter space it is applied to. This method is an extension of the one presented in~\cite{Gunther:2023xhh}, which was used in~\cite{Khalife:2023qbu} as a proof of concept for its functionality. Our method allows to fit any combination of data, since the training is performed specifically for the observables relevant to a given data set, while the emulator accuracy is matched automatically to a fraction of the observational errors. This approach works extremely well to speed up inference pipelines relying on many calls to already relatively fast theory codes, such as the Cosmic Linear Anisotropy Solving System, \class{}~\cite{Blas:2011rf,Lesgourgues:2011rh}, or the Code for Anisotropies in the Microwave Background, \camb{}~\cite{Lewis:1999bs,Howlett:2012mh}. Our emulator is called \ole{}\footnote{Available at \url{https://github.com/svenguenther/OLE}, with a detailed documentation available at \url{https://ole.readthedocs.io}.}, standing for \texttt{O}nline \texttt{L}earning \texttt{É}mulator. In this work, we describe the design of \ole{}, and showcase its applicability to inference using different cosmological data sets.

The paper is structured as follows. In Section~\ref{sec:ActiveEmu} we describe the emulator and its design. In Section~\ref{sec:examples}, we present a few examples of use cases of the emulator, and we compare its results and performance to traditional methods. We develop and benchmark a gradient-based staged sampling approach in Section~\ref{sec:diff_like}. We present our conclusions in Section~\ref{sec:conclusion}. In appendices~\ref{sec:usingOLE}~and~\ref{sec:precsions}, we provide more details on the code and how to use it.

\section{\ole{}{} Strategy}
\label{sec:ActiveEmu}
Emulators find widespread use in research that relies on computationally expensive simulations to perform parameter inference. Applications in cosmology are particularly common for N-body codes \cite{Heitmann:2008eq,Heitmann:2009cu,Lawrence:2009uk,Bhattacharya:2011vr,Agarwal:2012ew,Kwan:2012nd,Kwan:2013jva,Agarwal:2013aea,Heitmann:2013bra,Heitmann:2015xma,Lawrence:2017ost,Euclid:2018mlb,DeRose:2018xdj,McClintock:2018uyf,Zhai:2018plk,McClintock:2019sfj,Euclid:2020rfv,Bocquet:2020tes,Ho:2021tem,Moran:2022iwe,Jamieson:2022lqc,Kwan:2023yph,Upadhye:2023bgx, Brown:2024tom, Yang:2025dtc}, hydrodynamic simulations \nocite{adams1995hitchhikers}\cite{Bird:2018efe,Arico:2020lhq,Pedersen:2020kaw,Kugel:2023wte,Schaller:2024jiq,Walther:2024tcj} or emulators of Einstein-Boltzmann-Solvers \cite{Kaplinghat:2002mh,Jimenez:2004ct,Fendt:2006uh,Albers:2019rzt,Arico:2021izc,Gunther:2022pto,Nygaard:2022wri, bonici24, cosmopower22, Piras:2023aub, jense25, Bonici:2025ltp}. Beyond the architecture—such as neural networks or other machine learning techniques—one of the key design choices for an emulator is the distribution of support points used for training. A larger number of support points can enhance accuracy and enable reliable predictions across a broader parameter space. However, this comes at a significant computational cost, affecting both training and emulation, particularly in high-dimensional parameter spaces.

One way to reduce the number of training samples is through \textit{active sampling}, where the selection of additional support points is optimized based on the emulator's current state. This approach aims to improve accuracy in regions with high prediction uncertainty and relevance to the specific task the emulator is designed for.  
In the context of parameter inference, new support points should be concentrated in regions where high likelihoods are possible, but the emulator's accuracy remains insufficient.   
As the relevant parameter space only becomes available during the inference run itself, it comes natural to use \textit{online learning}, thus training the emulator during the inference on the samples it has seen before to have a suited emulator for all future samples to come.

\ole{} makes use of both active sampling and online learning, with the aim of deploying fast and compact emulators that rely on small but \textit{important} sets of samples.
Similarly to \textsc{connect}~\cite{Nygaard:2022wri}, \ole{} utilizes the sequential data of the inference process to train the emulator. This ensures that the collected samples are already concentrated in the most relevant regions of the parameter space. 
Unlike \textsc{connect}, \ole{} includes a built-in accuracy check, which can be applied to every individual call. This allows \ole{} to dynamically switch between the emulator and the standard theory code, to ensure accurate results and improve itself on-the-fly by adding support points whenever it is deemed necessary. To the best of our knowledge, this is the first emulator for cosmology using such an approach.

This strategy allows for an extremely data efficient construction of emulators that are trained on a set of support points which is a magnitude smaller than for comparable emulators.\footnote{For example, the \ole{} run outlined in Section \ref{sec:extended} on an extended set of cosmological parameters required 940 calls to \class{}, of which 240 were stored in the cache for training. Comparable analyses with \textsc{connect} used $\sim40$ times as many \class{} calls~\cite{Nygaard:2022wri}.} \ole{} requires no pre-training, as all data acquisition and training is done during the inference process itself.

\subsection{Learning Strategy}
\label{sec:learning_strategy}
In this Section, we outline the main features of the \ole{} algorithm, which is summarized via pseudocode in Algorithm~\ref{alg:ole}. 
We consider $N$ ordered samples of an arbitrary sampling algorithm.
For each sample $i$, we have a set of parameter values ${\boldsymbol\theta}_i$, observables ${\boldsymbol x}_i$, and a likelihood value $l_i$, such that a complete chain is:

\begin{align}
    \vec{X} &= ({\boldsymbol x}_1,...,{\boldsymbol x}_N)~, \\
    \vec{\Theta} &= ({\boldsymbol\theta}_1,...,{\boldsymbol\theta}_N)~, \\
    \vec{L} &= (l_1,...,l_N)~. 
\end{align}
For example, in a cosmological context, ${\boldsymbol\theta}_i$ would correspond to cosmological parameters, while ${\boldsymbol x}_i$ could represent the power spectra of the Cosmic Microwave Background (CMB). The parameters, observables, and likelihood values are related to each other via a theory code $\mathcal{T}$ and a likelihood code $\mathcal{L}$, 
\begin{align}
    {\boldsymbol x}_i &= \mathcal{T}({\boldsymbol\theta}_i)~, \\
    l_i &= \mathcal{L}({\boldsymbol x}_i)~.
\end{align}
The emulator $\mathcal{E}$ can, once trained, alternatively perform the role of the theory code. We denote emulated quantities via $\tilde{\boldsymbol x}_i$
\begin{align}
    \tilde{\boldsymbol x}_i &= \mathcal{E}({\boldsymbol\theta}_i)~.
\end{align}
\ole{} also defines an error estimation function $\texttt{IsEmulatorAccurate}({\boldsymbol \theta}_i)$ in charge of testing whether relying on the emulator is a valid choice for a set of parameters ${\boldsymbol \theta}_i$. This function estimates a statistic $\triangle_{\log l}$ that is proportional to the relative error on the likelihood induced by the use of the emulator. It returns a boolean flag, stating whether the emulator is acceptable or not on the basis of both $\triangle_{\log l}$ and $l_i$ itself, requiring more accuracy in the region where the likelihood is high.
Further details on the construction of this function and criterion can be found in Section~\ref{sec:unc_qualification}.

The algorithm begins by initializing the \ole{} internal \textit{cache}, which will store the parameters ${\boldsymbol\theta}_j$, the observables ${\boldsymbol x}_j$, and the corresponding likelihoods $l_j$. The process of filling the cache occurs in two stages.  
In the first stage, \ole{} populates the cache with all points generated by the sampler that fall within a specified margin of the highest likelihood found so far. 

During this \textit{burn-in} stage, the margin\footnote{This margin is computed from quantile function of the $\chi^2$-distribution with the argument of the probability $p$ and the dimensionality of the inference task. Thus, we would expect the occurrence of random samples of the posterior within the margin to happen with a likelihood of $1-p$. The probability $p$ is inferred from user-defined precision parameters, see Appendix~\ref{sec:precsions}.}
discriminates between relevant samples and outliers that are removed as they divert the focus of the emulator from the relevant parameter space.
Once a sufficient number of support points have been collected, the emulator is trained for the first time.

In the second stage, for each new sample ${\boldsymbol\theta}_i$, the emulator is evaluated and tested using the function $\mathtt{IsEmulatorAccurate}({\boldsymbol\theta}_i)$ defined above. If its accuracy flag returns $\mathtt{true}$, the emulated result $\tilde{\boldsymbol x}_i$ is used to compute the likelihood, and no additional training data is required.  
In contrast, if the emulator fails the accuracy test, this means that the emulator lacks precision in this region of parameter space. In that case, \ole{} calls the theory code $\mathcal{T}$ to recompute the observables at this point instead, potentially adding the new sample as a support point to the cache and updating the emulator.
This ensures good emulator performance for subsequent samples in this part of the parameter space.  
This approach guarantees that the emulator selectively adds points only in regions where its accuracy is insufficient, maintaining a low-correlation training set and thereby optimizing the learning process.

Ultimately, the emulator becomes accurate across the full region where the likelihood is high. Then, the function $\mathtt{IsEmulatorAccurate}({\boldsymbol\theta}_i)$ returns a positive result at each new point drawn from the sampler. In this final stage, the theory code no longer needs to be called, and the emulator greatly speeds up the sampling rate for the remainder of the run.

\begin{algorithm}
\caption{\ole{} sampling algorithm}
\begin{algorithmic}[1]
    \Require (\algorithmictype{functions}) {\tt Sampler,  FindLikelihoodThreshold, IsEmulatorAccurate}
    \Require (\algorithmictype{\ole{} emulator}) $\mathcal{E}$
    \Require (\algorithmictype{theory}) $\mathcal{T}$
    \Require (\algorithmictype{likelihood}) $\mathcal{L}$
    \Require (\algorithmictype{int}) $min\_data\_points$ (\ole{} hyperparameter), $N_\mathrm{steps}$ (number of sampler steps)
    \State ${cache} \gets$ \texttt{list}()
    \For{$i\gets 1, N_\mathrm{steps}$}  
        \State $\theta_i \gets \mathtt{Sampler}()$
        \If{\texttt{size}($cache$) $\le$ $min\_data\_points$}
            \State $x_i \gets \mathcal{T}(\theta_i)$
            \State $l_i \gets \mathcal{L}(x_i)$
            \State $ l_\mathrm{threshold} \gets \mathtt{FindLikelihoodThreshold(\textit{cache})}$
            \If{$l_i > l_\mathrm{threshold}$ }
                \State $cache$.\texttt{append}($x_i$)
            \EndIf
            \If{\texttt{size}($cache$) $=$ $min\_data\_points$}
                \State $\mathcal{E}$.\texttt{train}$(cache)$
            \EndIf
        \Else
            \If{$\mathtt{IsEmulatorAccurate}(\mathcal{E}, \theta_i)= \texttt{True}$}
                \State $\tilde{x}_i \gets \mathcal{E}(\theta_i)$
                \State $l_i \gets \mathcal{L}(\tilde{x}_i)$
            \Else 
                \State $x_i \gets \mathcal{T}(\theta_i)$
                \State $l_i \gets \mathcal{L}(x_i)$
                \If{$l_i > l_\mathrm{threshold}$ }
                    \State $cache$.\texttt{append}($x_i$)
                    \State $ l_\mathrm{threshold} \gets \mathtt{FindLikelihoodThreshold(\textit{cache})}$
                    \State $\mathcal{E}$.\texttt{train}$(cache)$
                \EndIf
            \EndIf
        \EndIf
        
    \EndFor
    \end{algorithmic}
\label{alg:ole}
\end{algorithm}

\subsection{Emulator Design}

For our active sampling strategy, we require an emulator that can estimate its own accuracy.
Similarly to \cite{Schneider:2010gv}, we find that \textit{Gaussian Processes} (GPs) applied to a compressed version of the data can achieve this goal, while reaching excellent accuracy even with a moderate training set size.

\subsubsection{Data Compression}

The observables we consider, such as the CMB angular power spectra and the matter power spectrum, are high-dimensional ($\mathcal{O}(10^3)-\mathcal{O}(10^4)$) but encode a relatively small number of features. To efficiently extract these features, we employ \textit{Principal Component Analysis} (PCA) for data compression (see Reference~\cite{bishop2007} for more details on PCA decomposition).

Our compression scheme transforms the data linearly, mapping it from the highly correlated and high-dimensional
\textit{data space} into the uncorrelated \textit{feature space} of PCA components. We find that between $10$ and $40$ orthogonal PCA components are usually sufficient to represent the cosmological observables considered in this work.  
More complex cosmological models, which involve a larger number of parameters, typically require more PCA components. In practice, the optimal number of components is determined in \ole{} based on its precision parameters, ensuring minimal information loss during the compression of observables. Since the features are computed from the data in the cache itself, with a growing cache the PCA components will eventually change. This results in a better resolution of subdominant physical effects and a better attribution of relevance between the found features. However, this change of PCA components is negligible for individual data points. Therefore, we keep the PCA components constant for most of the added points and then update them when a certain number of new samples is accumulated.\footnote{This strategy saves computational resources since a set of new PCA components requires an entire training of the emulator, while the addition of a single data point necessitates only a computationally cheaper update of the emulator.}
With \ole{}'s important attribute of removing outliers (see Section~\ref{sec:learning_strategy}), and since PCAs are sensitive to outliers, this strategy ensures that the compressed PCA components are the relevant features of the data. 

An ordinary PCA is agnostic to the experimental sensitivity and assigns equal weight to all features. However, since we have some prior knowledge of the experiment, we can optimize the compression by normalizing the data using the observational errors.\footnote{When no observational errors are provided to \ole{}, we assume them to be proportional to the sampled variance of the training set. This is equivalent to not prioritizing any specific part of the observable.}  
We find that this normalization reduces the number of required PCA components while improving the accuracy of the most significant features in the data.

\subsubsection{Gaussian Processes}

After transforming to feature space, we exploit the orthogonality property of the PCA components to construct a collection of independent one-dimensional emulators, each modeled using GPs, summarised in Appendix~\ref{app:GP}. This approach is highly beneficial since multidimensional GPs can be computationally expensive.\footnote{Multidimensional GPs scale quadratically with the emulated dimension because all cross-correlations must be inferred. In our case, since PCA components are approximately uncorrelated, the complexity is reduced from $\mathcal{O}(N^2)$ to $\mathcal{O}(N)$.}

The first few PCA components, representing the most prominent features of the data, often exhibit a relatively simple dependence on the input parameters ${\boldsymbol \theta}=\{\theta_1, ..., \theta_d\}$, typically following linear or quadratic relationships. In contrast, the high-order PCA components tend to encode a mixture of subdominant physical effects or even numerical noise induced by the theory code,
leading to more complex or even random dependency on the input parameters.

Every Gaussian Process relies on a choice of kernel and mean function, which must be optimized to accurately capture the response of a given PCA component to each parameter in the \( \boldsymbol{\theta} \) basis.  
To simplify the model, we normalize the data such that a zero (vanishing) mean function can be assumed—an approach that is commonly adopted when working with Gaussian Processes.
Hence, we select the prior of the Gaussian Process to be zero. Thus, without any training data the GP will return zero with a variance of one.

To efficiently describe the dependence of the PCA components on the input parameters, we employ a composite kernel comprising the sum of three components: an \textit{anisotropic radial basis function} (RBF) kernel, a \textit{linear} kernel, and a \textit{white-noise} kernel. This combination enables the emulator to better distinguish between physically meaningful variations and numerical noise, improving the robustness of the model.

The anisotropic RBF kernel allows the emulator to describe smooth variations of the PCA components in the vicinity of their distribution. Its hyperparameters consist of one length scale per input dimension (the dimensionality being $d$, the number of cosmological parameters). These length scales are fitted to the training set. They can be interpreted as the range over which a support point provides an accurate prediction, as neighboring points within this scale remain highly correlated. If no support point lies within this scale, the RBF kernel pushes the emulator to revert to the mean value of the PCA components while assigning a large emulation error. This typically prompts the addition of a new support point to the data cache, enhancing the emulation accuracy in future emulator calls.

The linear kernel instead captures the overall linear dependence of the PCA components on the input parameters and, in combination with the RBF kernel, enables the response of the PCA components to each parameter to be tilted overall. The linear kernel adds one hyperparameter per input dimension, which quantifies the amount of linear correlation that is fitted to the data. It is particularly beneficial for emulation far from the best-fit point in regions without nearby support points, where the linear kernel yields a suppression of the likelihood alongside a large prediction error. By incorporating the linear kernel, we can effectively disregard points that are too distant from the best-fit, improving the robustness of the emulator and preventing unnecessary additions to the support points far from the central emulation region, as long as the prediction error remains consistent with our precision goals. 

The white noise kernel adds one single hyperparameter accounting for the level of added white noise. We do not fit it to the data, but fix it according to the information loss incurred during data compression.\footnote{We compute the variance of the residuals introduced by the truncation of PCA components above a given order. We use this variance to estimate the overall white noise level. We distribute this white noise evenly between all PCA components, such that its impact is negligible for the dominant PCA components and significant for the noise-dominated ones. } This ensures that the inclusion of the noise kernel does not substantially inflate our overall noise budget. By setting an upper bound on the achievable accuracy of the GPs, the noise kernel prevents them from overfitting minor artifacts that have negligible practical impact on the final results. This leads to a more robust and efficient hyperparameter estimation. By refining the precision settings of the data compression process, this source of uncertainty is also automatically reduced.

\subsubsection{\ole{} Accuracy Estimation}
\label{sec:unc_qualification}

GPs offer a significant advantage in that they not only predict the expected value but also quantify their own uncertainty. This capability allows us to systematically assess whether the emulator is reliable within a given region of parameter space or if additional training points are still required.
When assessing the error of an emulated observable $\tilde{\mathbf{x}}$ at a given sample ${\boldsymbol\theta}$, it is relevant to account for its location in parameter space and, consequently, its likelihood $l$. While an error of a few percent may be acceptable far away from the global maximum likelihood, higher accuracy is required close to the best-fit point. Therefore, rather than imposing a uniform accuracy requirement for the GPs, we adapt an error threshold based on the likelihood value. This approach provides a natural interpretation that directly links the emulator's accuracy to the precision demands of the parameter inference task, ensuring that higher accuracy is prioritized in regions where it has the most impact on the final results.

Furthermore, the required precision of an emulated observable depends on its inherent measurement accuracy. An observable measurable at per-mille precision imposes stricter emulation accuracy requirements compared to one that is only determined at order unity. 

\begin{algorithm}
\caption{\ole{} algorithm for the \texttt{IsEmulatorAccurate} function.}
\textbf{Input:} (\algorithmictype{sample}) $\theta$ \\
\textbf{Output:} (\algorithmictype{bool}) \textit{accurate\_flag} 
\begin{algorithmic}[1]
    \Require (\algorithmictype{\ole{} emulator}) $\mathcal{E}$
    \Require (\algorithmictype{likelihood}) $\mathcal{L}$
    \Require (\algorithmictype{function}) \texttt{PrecisionCriterion}
    \Require (\algorithmictype{int}) $N\_quality\_samples$ (\ole{} hyperparameter)
    \State $l_\mathrm{samples}$ $\gets $ \texttt{list}()
    \For{$i \gets 1$, $N\_quality\_samples$} 
        \State $\tilde{x}_\mathrm{PCA, sample} \gets \mathcal{E}.\texttt{sample\_from\_GP}(\theta)$
        \State $\tilde{x}_\mathrm{sample} \gets \mathcal{E}.\texttt{decompress}(\tilde{x}_\mathrm{PCA, sample})$      
        \State $\tilde{l}_\mathrm{samples}$.\texttt{append}($\mathcal{L}(\tilde{x}_\mathrm{sample})$)
    \EndFor
    \State $\tilde{x} \gets \mathcal{E}(\theta)$
    \State $\tilde{l} \gets \mathcal{L}(\tilde{x})$
    \State $\triangle_{\log \tilde{l}} \gets \sqrt{\texttt{median}((\log \tilde{l}_\mathrm{samples}-\log \tilde{l})^2)}$
    \State $\triangle_{\log \tilde{l},\mathrm{max}} \gets $\texttt{PrecisionCriterion}($\log \tilde{l}$)
    \If{$\triangle_{\log \tilde{l}} < \triangle_{\log \tilde{l},\mathrm{max}}$}
        \State \textit{accurate\_flag} $\gets$ \texttt{True}
    \Else
        \State \textit{accurate\_flag} $\gets$ \texttt{False}
    \EndIf 
\end{algorithmic}
\label{alg:testing}
\end{algorithm}
Eventually, we estimate an error on the likelihood by sampling from the GP emulator and computing the response of the likelihood function (see the pseudocode of the \texttt{IsEmulatorAccurate} Algorithm~\ref{alg:testing}). This procedure samples a set of observables from the error estimate of the GPs that are computed for a given point ${\boldsymbol\theta}_i$. By evaluating the likelihood for each of these plausible observables, we obtain a distribution of likelihood values, which allows us to quantify the uncertainty of the likelihood estimation.

Instead of directly measuring the standard deviation from this collection, which is susceptible to large sample variance, we construct a more robust statistic by sampling the GPs at their one-sigma boundary. We then compare the median difference of these samples to the likelihood obtained from the mean prediction of the emulator. This approach provides a more stable and reliable measure of the uncertainty in the likelihood estimation, reducing sensitivity to outliers and improving the robustness of our accuracy assessment.

In order to combine the errors of our individual GP emulators into one final error for the likelihood,
we assume that all emulators are entirely uncorrelated, despite relying on the same support points, and that they contribute independently to the likelihood evaluation. In practice, these assumptions are often not fulfilled, leading to an overly optimistic estimation of the uncertainty. 
Nevertheless, these effects primarily introduce a constant scaling factor in our error estimates. This scaling factor can be empirically determined during the deployment of the emulator by comparing the estimated errors with the direct results obtained from the theory code. This calibration step allows us to correct for the overestimation of precision and improve the reliability of our uncertainty quantification.

In the following we look at examples of the scaled error estimate. In Figure~\ref{fig:error_statistic} we compare the difference between the (log-)likelihood of samples that were computed using the emulator ($\tilde{l}$) and the (log-)likelihood of samples that used the theory code ($l$). All samples were randomly selected from the Markov chain Monte Carlo (MCMC) analysis of Example 1 in Section \ref{sec:lcdm}. In Figure~\ref{fig:h_vs_loglike}, one observes the distribution of the samples' likelihood and the distribution of the Hubble parameter. The largest likelihood can be obtained in the center of the Hubble parameter distribution. This supports our expectation that the best-fit point is located near the center of our MCMC samples of single-peaked, (approximately) symmetric distributions. 
The difference between the simulation-inferred likelihood and the emulated one is encoded by the color. The largest deviations can be found at the smallest likelihoods, where their influence is smaller. Overall, we observe no systematic over- or underestimation which would directly lead to systematic biases in the estimate of the posterior. In Figure~\ref{fig:devs_hist} we show a histogram of the differences in the (log-)likelihoods scaled by the error estimator. As before, the histogram is centered at zero which indicates that there is no overall bias. The standard deviation is about unity, which indicates that the error estimate is accurate.\footnote{We find that the distribution is not necessarily Gaussian but exhibits wider tails on both sides. However, this does not change our interpretation.}

Having validated the statistic of the error estimate of \ole{}, we still need to construct a validity criterion that dictates whether the \ole{} prediction, given its uncertainty, will or will not be used. In Algorithm \ref{alg:ole}, we denote this function \texttt{IsEmulatorAccurate}. We choose a validity criterion such that the error estimate $\triangle_{\log \tilde{l}}$ must be smaller than the maximum acceptable error $\triangle_{\log \tilde{l},\mathrm{max}}$ for the accuracy to be deemed acceptable. This upper bound consists of three terms
\begin{align}
    \triangle_{\log \tilde{l},\mathrm{max}} = c_0 + c_1 \log \frac{\tilde{l}}{l_\mathrm{bf}} + c_2 \log^2 \frac{\tilde{l}}{l_\mathrm{bf}},
    \label{eq:prec}
\end{align}
that depend on the three user-given precision parameters $c_i$ for $i\in(0,1,2)$, the estimated likelihood of the emulator $\tilde{l}$ and the likelihood of the current best-fit that was found by the sampler $l_\mathrm{bf}$. The first parameter $c_0$ represents the acceptable ground level error around the maximum of the distribution. Note that this parameter should be chosen as a function of the dimension of the sampled distribution\footnote{An error of $0.1$ log-likelihood is very relevant for a 1 dimensional distribution while it might not be for a higher dimensional one. For example, the difference between the best-fit and the 1$\sigma$ bound for 1-dimensional case is at $0.5$ log-likelihood for the 1-dimensional case, while in 27 dimensions it is at $\sim14.9$.}. The other two terms serve to degrade the required accuracy at the edges of the distribution. 

For Gaussian posterior distributions, the linear precision parameter $c_1$ has an intuitive interpretation.
In this case, the log-likelihood is quadratic in the sampled parameters $\theta$. Hence, when inferring the relation between the log-likelihood and $\theta$, the first derivative is linear in $\theta$. Thus, the credible intervals in $\theta$ (e.g. around the 1$\sigma$ boundary) are related to credible intervals in $\log l/l_\mathrm{bf}$, such that a change in $\log l/l_\mathrm{bf}$ will be directly proportional to a change in $\theta$. For the relative errors, one finds that $\frac{\triangle \log l}{\log l/l_\mathrm{bf}} = 2 \frac{\triangle \theta}{\theta}$. In the case where Equation~\ref{eq:prec} is dominated by the linear term one can directly relate $c_1$ to $2\frac{\triangle \theta}{\theta}$. Thus, in order to determine the position of the $1\sigma$ contour to a percent level, one should set $c_1 = 0.02$.

The quadratic term can be seen as a cutoff scale at which the demand on accuracy falls off. The default value of this term is set such that it dominates Equation~\ref{eq:prec} at log-likelihoods outside of the $3-4\ \sigma$ parameter region at which little calls are expected anyways.

\begin{figure}
      \begin{subfigure}{0.49\textwidth}
        \includegraphics[width=\linewidth]{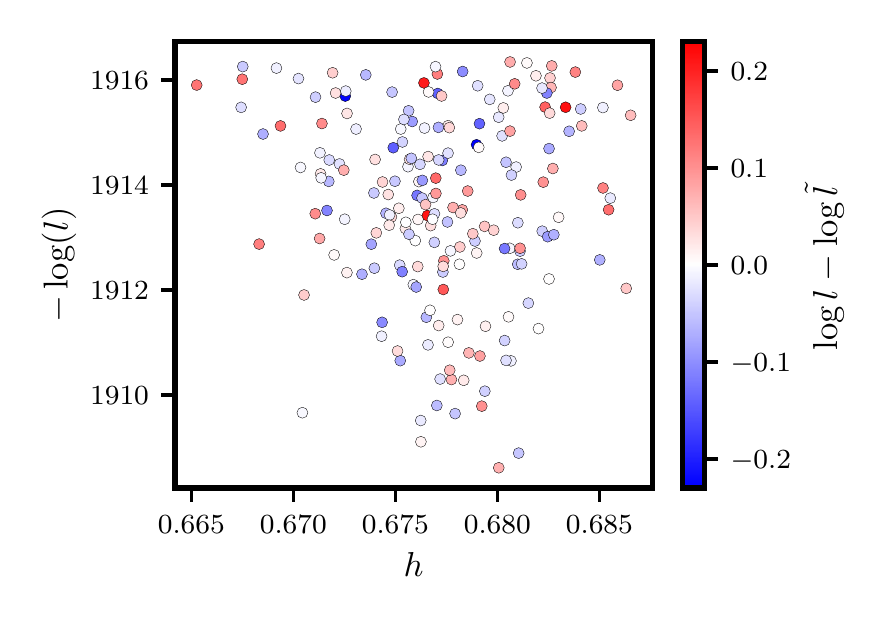}
        \caption{} \label{fig:h_vs_loglike}
      \end{subfigure}%
      \hspace*{\fill}   
      \begin{subfigure}{0.49\textwidth}
        \includegraphics[width=\linewidth]{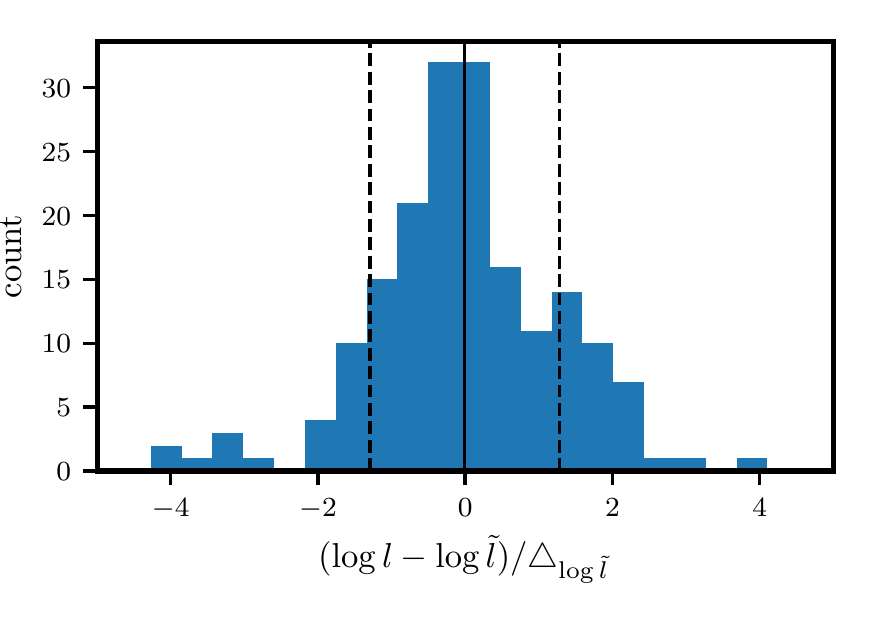}
        \caption{} \label{fig:devs_hist}
      \end{subfigure}%
    \caption{Evaluation of the log-likelihoods computed with predictions from \class{} ($\log l$) and \ole{} ($\log \tilde{l}$) in the example of Section \ref{sec:lcdm}. On the left they are displayed as function of the Hubble rate $h$. The over or undershooting of the estimated likelihood from using the \ole{} prediction is indicated by color. On the right the difference between the likelihoods are scaled by the error estimate of the emulator $\triangle_{\log \tilde{l}}$. The solid and dashed black lines indicate the one standard deviation region and mean, respectively. The likelihood values calculated using the emulator prediction are unbiased and the emulator's uncertainty estimation is accurate.}
    \label{fig:error_statistic}
\end{figure}

\subsection{Code Design}

The code can be interfaced with the cosmological samplers \montepython{}\footnote{\url{https://github.com/brinckmann/montepython_public}} \cite{Brinckmann:2018cvx} and \cobaya{}\footnote{\url{https://github.com/CobayaSampler/cobaya}} \cite{Torrado:2020dgo}. Therefore, \ole{} is designed to be parallelized with MPI by constructing an emulator for each MPI-process, each corresponding to an MCMC chain. In order to increase efficiency, all chains share one cache such that simulation calls can be shared among all processes. This significantly reduces the duration of the initial data collection to train the emulator. Furthermore, the training of the emulator is parallelized and MPI communication is implemented such that updates to one process are forwarded to all other processes, making all emulators based upon the same set of support points. 

A major acceleration of the emulator can be achieved by using the \texttt{JAX} software package \cite{jax2018github}, which is specialized on high-performance numerical computing. The \textit{just-in-time} (\textit{jit)} compilation of \texttt{JAX} transforms the functions of the emulator and compiles it during the runtime of the code. This enables the emulator to be executed efficiently, accelerating it by a factor of $\sim\mathcal{O}(10^2)$. However, this compilation also results in an increase in memory consumption compared to using just the standard theory code.\footnote{An example MCMC with 12 parallel MPI processes requires $\sim5$ GB of RAM with standard \class{} and 35-80 GB with \ole{}, depending on precision settings. The biggest memory requirement comes from the `jitted' emulator, as each MPI process needs to keep this in its memory. The more points in the cache, the more memory each process will use. The total memory usage can thus be reduced by running fewer parallel processes, but in practice this should rarely be a problem on a modern computing cluster.}

\texttt{JAX} also exposes \ole{} to an automatic differentiation algorithm.
This makes it easy to obtain robust derivatives of cosmological observables with respect to parameters.
Using a series of examples, References \cite{campagne23, Balkenhol:2024sbv, piras24} illustrate how access to such derivatives facilitates faster and more efficient analysis pipelines;
we add to this with the development of a staged gradient-based sampler in Section~\ref{sec:diff_like}.

Lastly, for the implementation of the GPs, we use the python package \texttt{GPJax}\cite{Pinder2022} that comes with an efficient \texttt{JAX}-compatible implementation for the construction, training and evaluation of GPs.

\subsection{\ole{} in MCMC Analyses}
\label{sec:OLE_mcmc}
Much of modern cosmological inference is done via MCMC analyses with Metropolis Hastings (MH) sampling, which requires a large number of calls to a cosmological theory code to compute the observables for a given set of input parameters. Using \ole{} greatly reduces the time spent on this aspect, speeding up the acquisition of samples, eventually reaching a level of speed-up where the runtime of the likelihood code is the dominant step.

During the initial burn-in phase, the total runtime is dominated by evaluating the theory code in order to burn-in and accumulate training data. In this phase, we use oversampling (keeping the `slow' cosmological parameters fixed during several steps while sampling over `fast' nuisance parameters\footnote{See Reference~\cite{Torrado:2020dgo} for details.}) to converge more quickly to the center of the posterior distribution. Once the training is done, \ole{} proceeds with the test phase in which the execution time is largely taken up with testing/training and updating the emulator. However, the testing rate is one of the limiting factors in the achievable acceleration of the MCMC. As this necessitates multiple calls to the likelihood code, too frequent testing can quickly become a computational bottleneck. Therefore, the emulator gradually lowers the testing rate until it reaches a user-defined minimum. Eventually, once the emulator is confident and the testing rate is reduced, the runtime is dominated by the evaluation of the likelihood rather than the theory code evaluation. When \ole{} reaches a level of accuracy where only a negligible part of the total MCMC runtime is spent evaluating the theory code, we can derive an upper bound on the total acceleration induced by the emulator. Assuming that the evaluation time of the emulator ($T_\mathrm{\mathcal{E}}$) is much smaller than the one of the theory ($T_\mathrm{\mathcal{T}}$) and likelihood ($T_\mathrm{\mathcal{L}}$) codes, we get:

\begin{equation}
    \textit{Speed-up} = \frac{T_\mathrm{\mathcal{T}} + T_\mathrm{\mathcal{L}}}{T_\mathrm{\mathcal{L}}}~.
    \label{eq:speedup}
\end{equation}

This speed-up is generally achievable for a variety of likelihood codes and cosmologies, as we present here, and is particularly evident in Figure~\ref{fig:lcdm_timings}.
Another consequence of the likelihood code dominating the runtime is that oversampling is no longer useful. It even slows down the acquisition rate of points with new cosmological parameters in the chains. Oversampling is only beneficial when the emulator remains slower than the likelihood code or during the initial burn-in phase, when the emulator is rarely used. Hence, in this stage of the MCMC run, \ole{} automatically switches off the fast/slow parameter decomposition.

The progress of an example chain, and with it of the emulator, is displayed in Figure~\ref{fig:lcdm_h}, using a chain from our standard $\Lambda$CDM example (see Section~\ref{sec:lcdm}). We distinguish between the initial burn-in phase of about 300 calls to \class{} during which the chain moves towards the center of the posterior and the acquisition of the first data points used to train the emulator. While about 2/3 of those calls are regarded as outliers of little use to the emulator in the latter stage (which is dominated by calls in the most central region of the parameter space), about 1/3 of the initial calls remain in the training set. Once $\sim 80$ training points are gathered, the emulator is trained for the first time. In the early phase after the initial training (and in particular after new points are added to the cache), most emulator calls are tested to ensure a sufficient accuracy of the emulated quantities. During this phase, ``holes'' in the training set that lead to inaccurate predictions are detected and filled by additional calls to the theory code. Eventually, most gaps are closed and the emulator is confident with its predictions (achieving the user-defined desired accuracy level). 

The most important \ole{} precision parameters are presented in Appendix~\ref{sec:precisions_default}, along with their default values.

\begin{figure}
    \centering
    \includegraphics[width=0.99\linewidth]{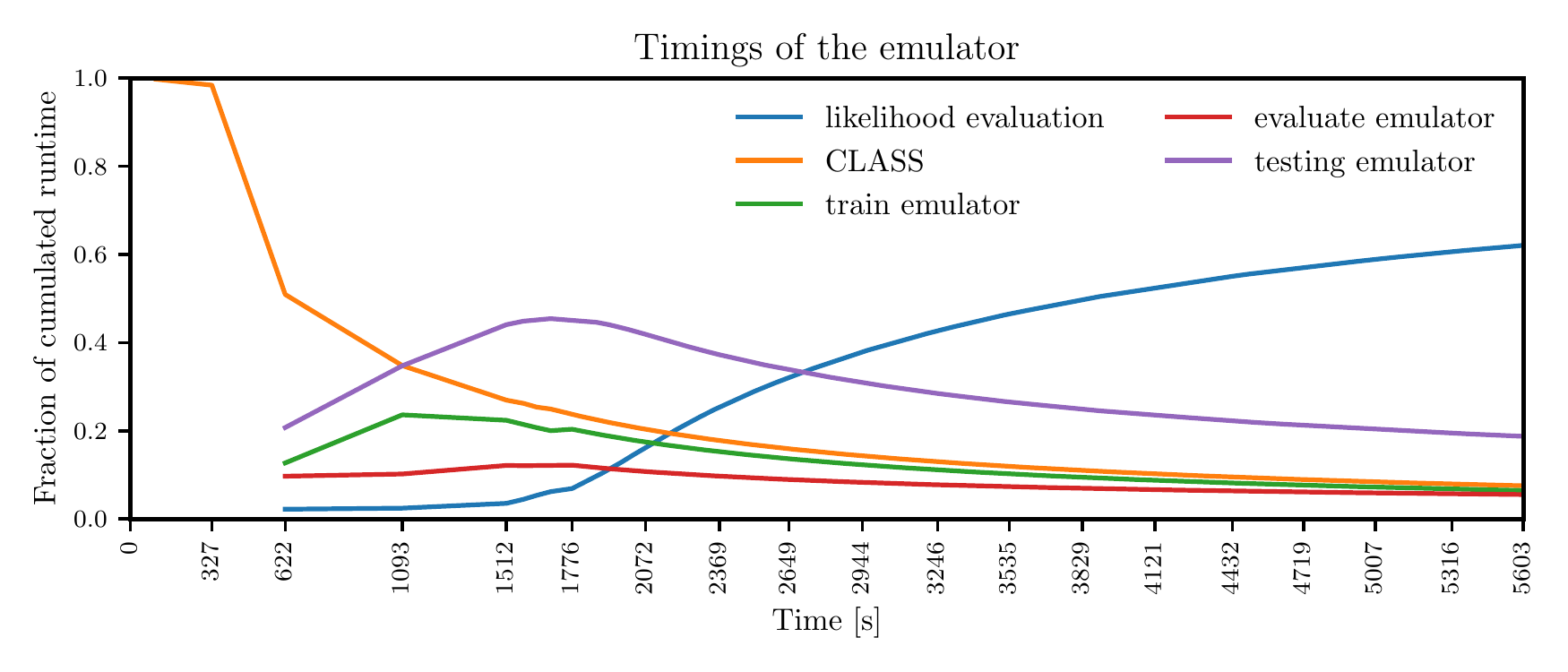}
    \caption{Fraction of elapsed computation time spent on each component of the process, spanning the first $\sim1.5$ hours of an MCMC analysis with \ole{}. It is clear that the initial phase is dominated by running the traditional theory code until enough points have been computed to train the emulator. Then comes a phase dominated by testing the emulator, as its accuracy must be verified. Finally, when confidence in the emulator is high, the runtime becomes dominated by the likelihood. As the run progresses, it will become even more likelihood-dominated.}
    \label{fig:lcdm_timings}
\end{figure}

\begin{figure}
    \centering
    \includegraphics[width=0.99\linewidth]{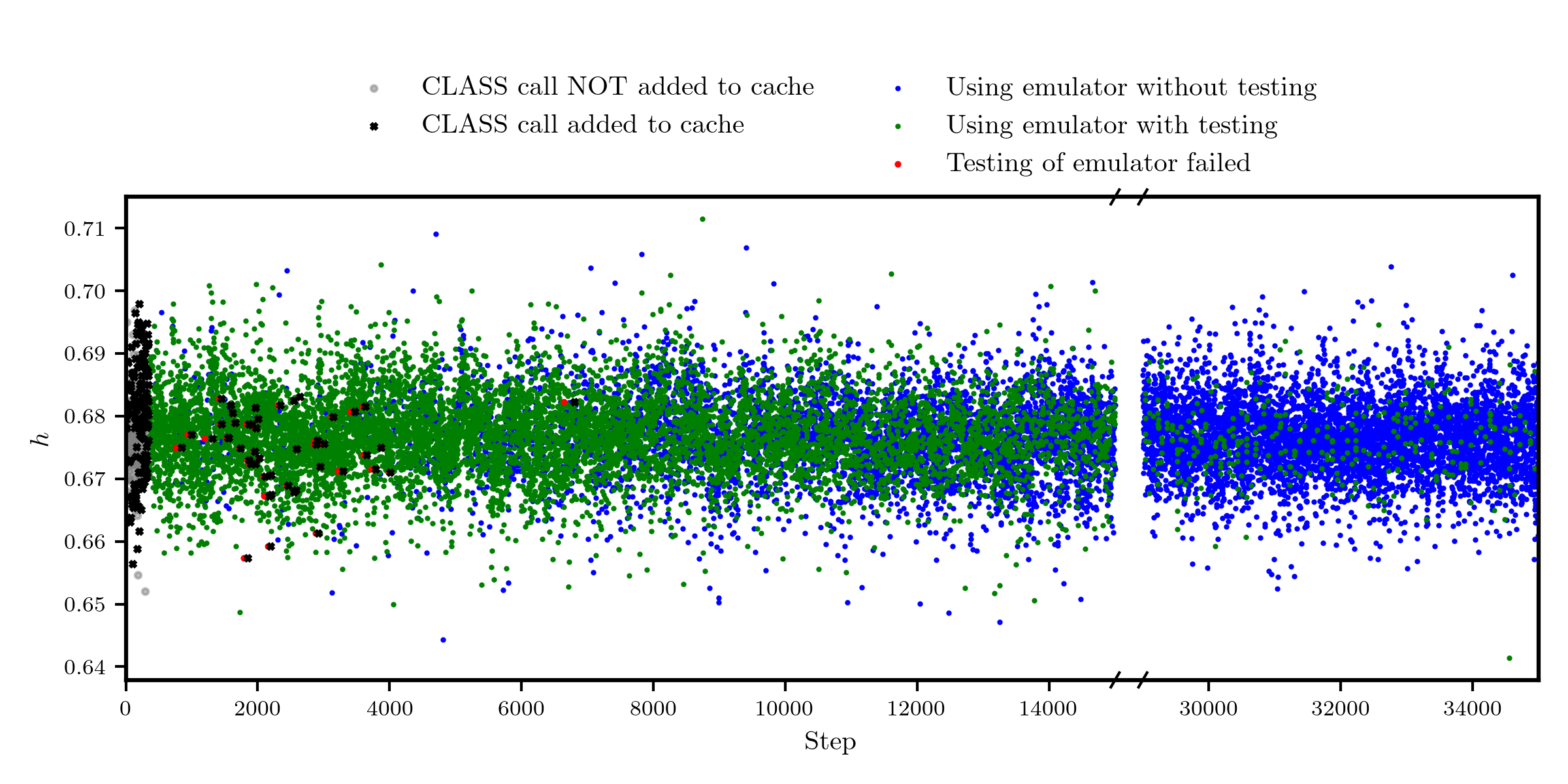}
    \caption{
    Value of the reduced Hubble parameter $h$ for the first 15000 theory evaluations in one of the chains of a \cobaya{} MCMC run, as well as 6000 evaluations later in the chain for comparison. The points are colored according to the type of call to the standard theory code or emulator: \class{} calls saved in the cache and used to improve the emulator (black cross), \class{} calls deemed not relevant for the emulator and therefore not cached (gray dots), emulator calls followed by a successful accuracy test (green dots), emulator calls followed by a failed accuracy test (red dots), and emulator calls not followed by an accuracy test (blue dots). Gray dots appear only at the beginning during the burn-in phase. Note that red dots (failed accuracy tests) are immediately followed by black dots (\class{} calls added to the cache). As performance checks are passed and confidence in the emulator grows, testing is gradually decreased, such that green dots give way to blue dots. Note that points very far from the high-likelihood region would always be  computed by the standard theory code to avoid an overly expensive and conservative training of the emulator, but this situation did not occur within the displayed sample.
    }
    \label{fig:lcdm_h}
\end{figure}

\section{Examples with Metropolis Hastings Sampling}
\label{sec:examples}

We demonstrate the accuracy and efficiency of \ole{} as a tool for parameter inference through a selection of examples covering a variety of samplers, theory codes, cosmological models and observational likelihoods. In this section, we limit ourselves to the use of the MH algorithm as implemented in the \cobaya{} or \montepython{} samplers, and to the theory codes \class{} and \camb{}. The next section features additional sampling algorithms and parameter inference packages.

\vspace{0.1cm}

\noindent {\it Quantifying efficiency.}
When running \cobaya{}, we sample until reaching a desired value of the convergence criterion,\footnote{We use the Gelman-Rubin statistic $|R-1|$~\cite{GelmanRubin}.} while with \montepython{} we run our chains until hitting a given number of points or a maximum wall-clock time. To quantify the efficiency of \ole{} versus traditional runs, we could simply compare the time it takes on a given platform to reach a given convergence level or a given number of points. Such a test would not be robust and informative enough, because such a time could be strongly affected by arbitrary choices for the number of chains running in parallel, the number of cores per chain, the oversampling factor for nuisance parameters, the initial proposal density, etc. In order to have a uniform metric, we refer to the effective sample size (ESS) of the cosmological parameters of MCMC chains (see Reference~\cite{brooks2011handbook}). The ESS provides an estimate of the effective number of uncorrelated samples in a chain. It is obtained by first estimating the scale on which the samples of a chain are correlated, and then dividing the number of total samples by this correlation scale.\footnote{In this work, we compute the ESS using \getdist{}~\cite{Lewis:2019xzd} and \arviz{}~\cite{arviz19}.} Note that oversampling allows to acquire samples at a higher rate in traditional runs at the expense of increasing the correlation scale, such that each drawn (cosmological) sample contains less information. Thus, using the ESS instead of the number of samples allows us to have a meaningful comparison between the performance of runs performed with or without oversampling. This is useful in our context since, as mentioned in Section~\ref{sec:OLE_mcmc}, \ole{] switches off oversampling when no longer useful. The CPU time needed on a given platform to reach a given ESS directly estimates the efficiency of a MH sampling pipeline, since it quantifies the speed at which relevant information accumulates. For the CPU time, we should use the effective amount of time during which the CPUs were active, which reflects the energy consumption of the runs. This time can be obtained by multiplying the wall-clock time of a run by the number of chains, the number of cores per chain, and the average CPU usage efficiency reported by the computer at the end of the run. We denote the effective CPU time in hours as CPUh. In summary, we will compare the efficiency of different runs performed with or without \ole{} by looking at the ratio of their ESS per effective CPU time in hours, ESS/CPUh.

\vspace{0.1cm}

\noindent{\it Quantifying accuracy.}
Various estimators can be defined to quantify the agreement between parameter inference runs based on \ole{} and traditional runs. For simplicity and concision, we will focus on the  shift $D_x$ of the mean value of the 1-dimensional posterior for each parameter $x$ normalized to the standard deviation found in the traditional run, $\sigma_x$. This dimensionless shift is given as
\begin{equation}
    D_x = \frac{\left|\left\langle x \right\rangle - \left\langle \tilde{x} \right\rangle\right|}{\sigma_x},
    \label{eq:disagreement}
\end{equation}
where $\left\langle \tilde{x} \right\rangle$ refers to the mean found in the \ole{}-based run, and $\left\langle x \right\rangle$ to that of the traditional run. Even when comparing two well-converged runs, sampling noise usually leads to shifts $D_x$ in the range from 0.01 to 0.1~\cite{Gunther:2022pto}, indicating that the mean value is estimated with good accuracy compared to the parameter uncertainty. In the following sections, we will quote the relative shifts $D_x$ obtained for various parameters, assuming different cosmological models and likelihoods. Values of $D_x$ smaller than 0.1 will indicate that the \ole{} emulator is not significantly biasing the posteriors.

\subsection{$\Lambda$CDM with Current Data Using \cobaya{} and \class{}}
\label{sec:lcdm}}

\noindent {\it Model.} As a first example, we investigate the performance of the emulator by running \cobaya{} for the standard $\Lambda$CDM model, both with and without the use of \ole{}. We parameterize the $\Lambda$CDM model via:
the expansion rate today $h$ in units of $100\,\mathrm{km/s/Mpc}$, the fractional density of baryons $\Omega_{\mathrm{b}}$ and cold dark matter $\Omega_{\mathrm{cdm}}$, the amplitude $\ln(10^{10} A_{\mathrm{s}})$ and spectral index $n_{\mathrm{s}}$ of the primordial power spectrum of scalar fluctuations, as well as the optical depth to reionization $\tau_\mathrm{reio}$. One of the three neutrino species is assumed to have a fixed mass of $0.06$~eV. A definition of all cosmological parameters referenced in this work can be found in Table~\ref{tab:par_def} of Appendix \ref{sec:par_def}.

\vspace{0.1cm}

\noindent {\it Data.} We include \planck{} 2018 data for temperature, polarization~\cite{Planck:2019nip} and lensing~\cite{Planck:2018lbu} anisotropies, as well as Supernova luminosity data from the Pantheon sample \cite{Pan-STARRS1:2017jku} 
and Baryon Acoustic Oscillation (BAO) data from the SDSS DR7~\cite{Ross:2014qpa} and DR12~\cite{BOSS:2016wmc} samples. In both types of runs, we use \class{} as the theory code to compute observables, and the MH sampler of \cobaya{} to run 8 parallel chains up to a Gelman-Rubin statistic of $|R-1|<0.01$. The likelihood codes used here are not as highly parallelized as the theory codes. Since the \ole{} runtime is dominated by likelihood evaluations, it is not extremely efficient to run chains on many cores in the \ole{} case. A higher number of cores would speed up the initial phase during which the theory code is often called, but then, once the run time gets dominated by non-parallelized-likelihood evaluations, it would result in an inefficient use of resources with no benefit. We thus run each of the 8 chains on 4 cores each in the traditional run, and on a single core each in the \ole{} run. We also mentioned in Section~\ref{sec:OLE_mcmc} that oversampling is no longer useful with \ole{}. In the traditional run, we use the default \cobaya{} settings for oversampling the 21 nuisance parameters, but in the \ole{} run, the fast/slow parameter decomposition gets switched off automatically after the initial training phase. We perform the \ole{} run using the default values of the \ole{} precision parameters described in Appendix~\ref{sec:precisions_default}. 

\vspace{0.1cm}

\noindent {\it Accuracy.} We find excellent agreement between the posteriors derived from the two runs, as can be seen in Figure~\ref{fig:lcdm}. The normalized shifts of the means $D_x$ of Equation~\eqref{eq:disagreement} are found to be below 0.03. Such tiny deviations are consistent with the stochastic variance of the posteriors estimated from MCMCs~\cite{Gunther:2022pto}.

\begin{figure}
    \centering
    \includegraphics[width=0.9\linewidth]{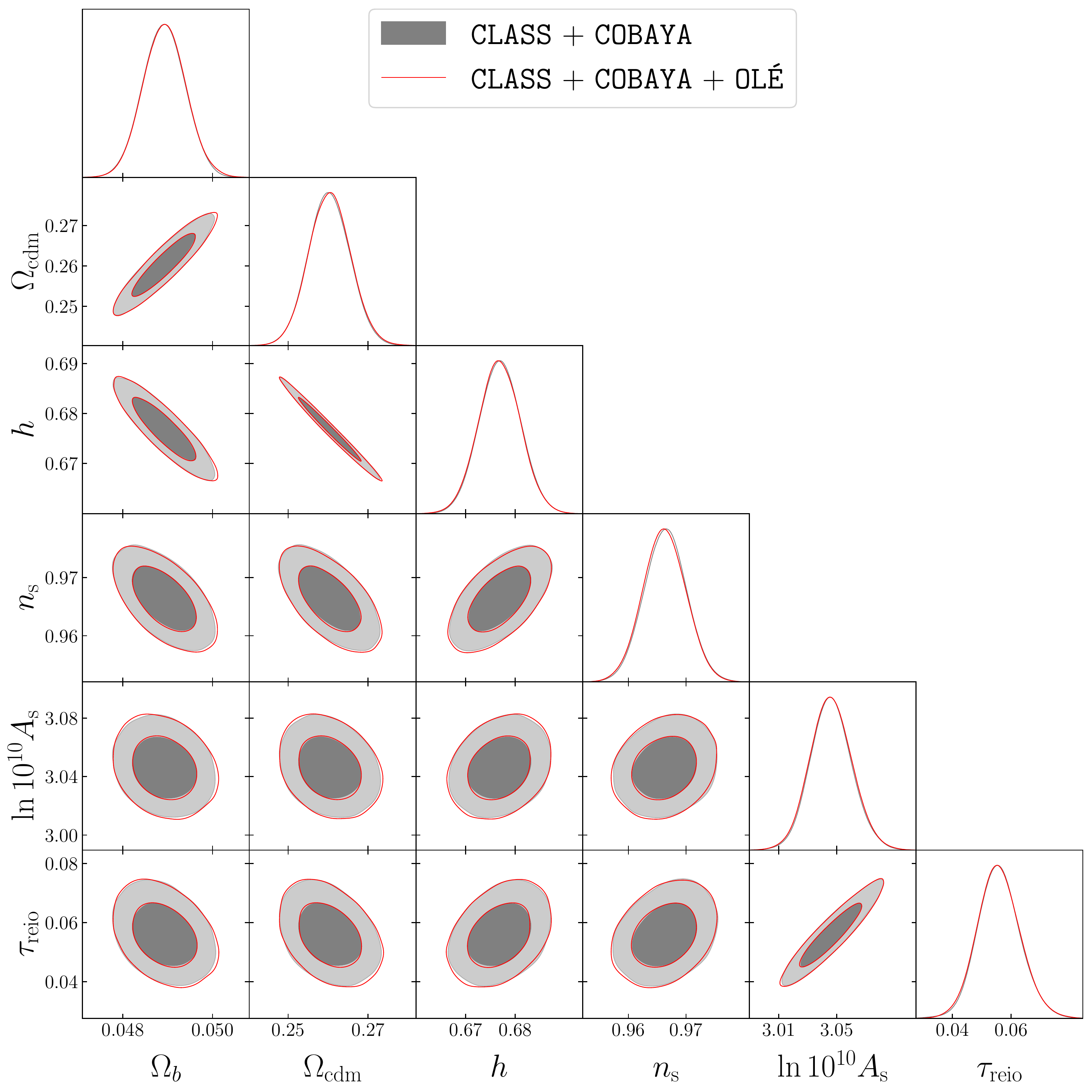}
    \caption{Posteriors of the $\Lambda$CDM model parameters inferred from \class{} and \cobaya{} with \planck{} 2018, Pantheon+ and BAO data, as outlined in Section~\ref{sec:lcdm}. We see an excellent match between the posteriors computed using either a standard MCMC or the \ole{} emulator, with a maximum deviation of posterior means by $0.03\sigma$.}
    \label{fig:lcdm}
\end{figure}

\vspace{0.1cm}

\noindent {\it Efficiency.} In order to estimate the maximum achievable speed-up according to Equation~\eqref{eq:speedup}, we measure $T_\mathcal{T} \sim 5 \, \mathrm{sec}$ and $T_\mathcal{L} \sim 0.05 \, \mathrm{sec}$ on 4 cores in this particular case. Consequently, we may expect an acceleration by at most about a factor of 100. Note that this is only a rough estimate, which does not take into account several effects such as the initial burn-in phase, the computation efficiency, the cost of calling \class{} to build the cache, that of testing the emulator accuracy, or the impact of oversampling over nuisance parameters.\footnote{Even if, as explained in Section~\ref{sec:examples}, using the ESS instead of the number of points in the chains mitigates the impact of using oversampling.} In practice, by comparing the sampling rate ESS/CPUh in the two runs, we find that \ole{} accelerates the whole parameter inference pipeline by a factor $\sim$98, which is very close to the expected speed-up.

\vspace{0.1cm}

\noindent We conclude that \ole{} is able to boost the efficiency of MCMC runs by a considerable amount without degrading their accuracy. However, it is worth checking that this conclusion applies also to more complex models and other likelihoods.

\subsection{Extended Cosmology with Current Data Using \cobaya{} and \camb{}}
\label{sec:extended}

\noindent {\it Model.} To investigate the performance of \ole{} in the case of computationally more expensive models, we extend the $\Lambda$CDM cosmology to include a non-zero spatial curvature parametrised with the effective density $\Omega_k$, a (degenerate) neutrino mass leading to the fractional neutrino density $\Omega_\nu$, and a dark energy (DE) fluid with an equation of state $w(a)$ obeying the Chevallier-Polarski-Linder (CPL) parametrization \cite{Linder:2002et, Chevallier:2000qy}
\begin{equation}
    w(a)=w_0 + w_a(1-a)~,
    \label{eq:CPL}
\end{equation}
with free parameters $w_0$ and $w_a$. For the minimal $\Lambda$CDM parameters, to illustrate the versatility of the code, we use the same parameters as in the previous subsection except for the total non-relativistic matter density $\Omega_\mathrm{m}$ (replacing $\Omega_\mathrm{cdm}$) and the current matter power spectrum amplitude parameter $\sigma_8$ (replacing $A_\mathrm{s}$). Note that for the \ole{} emulator, the effect of the neutrino mass (or equivalently of $\Omega_\nu$) on the CMB spectra is potentially more challenging to capture than that of other parameters. As a matter of fact, for very small values of $\Omega_\nu$, the response of the CMB spectrum coefficients $C_\ell$ to variations of $\Omega_\nu$ is far from linear.

\vspace{0.1cm}

\noindent {\it Data.} We base this analysis on \planck{} 2018 temperature, polarization and lensing measurements, in conjunction with DESI 2024 BAO data \cite{DESI:2024mwx}. The observables are computed with \camb{} using the Parametrized Post-Friedmann (PPF) DE parametrization \cite{Fang:2008sn}. We use again the MH sampler of \cobaya{}. As in the previous example, we carry out two runs, a traditional one with 8 chains using 4 cores each, and an \ole{} run with 8 chains using a single core each. Both runs proceed until reaching a Gelman-Rubin statistic of $|R-1|<0.02$. Like in the previous subsection, we oversample the 21 nuisance parameters only in the traditional run and in the initial training phase of the \ole{} run.
As explained in Appendix~\ref{sec:precisions_camb}, we perform the \ole{} run using default precision, except for two parameters. First, for a faster run, we loosen our requirement on the log-likelihood error $\Delta \ln {\cal L}$ below which the emulator is considered valid. However, to get an accurate emulation of the physical effect of the neutrino mass, we enhance the accuracy setting controlling the number of PCA components.

\vspace{0.1cm}

\noindent{\it Accuracy.} Figure~\ref{fig:extended} shows an extremely good level of agreement between the posterior distributions  obtained in the two cases. In particular, the non-Gaussian nature of the posteriors for additional parameters is well captured by both MCMCs. The normalized shifts of the means $D_x$ between the posteriors computed with or without \ole{} never exceeds 0.048, which is comparable to the results of the previous subsection. Note that the physical observables of the extended model feature four additional physical effects compared to the $\Lambda$CDM case. We find that the \ole{} emulator is able to capture these additional effects and to maintain the same final level of accuracy on parameter inference. This illustrates the robustness and the stability of the automatic error estimation algorithm implemented in \ole{}.

\begin{figure}
    \centering
    \includegraphics[width=0.95\linewidth]{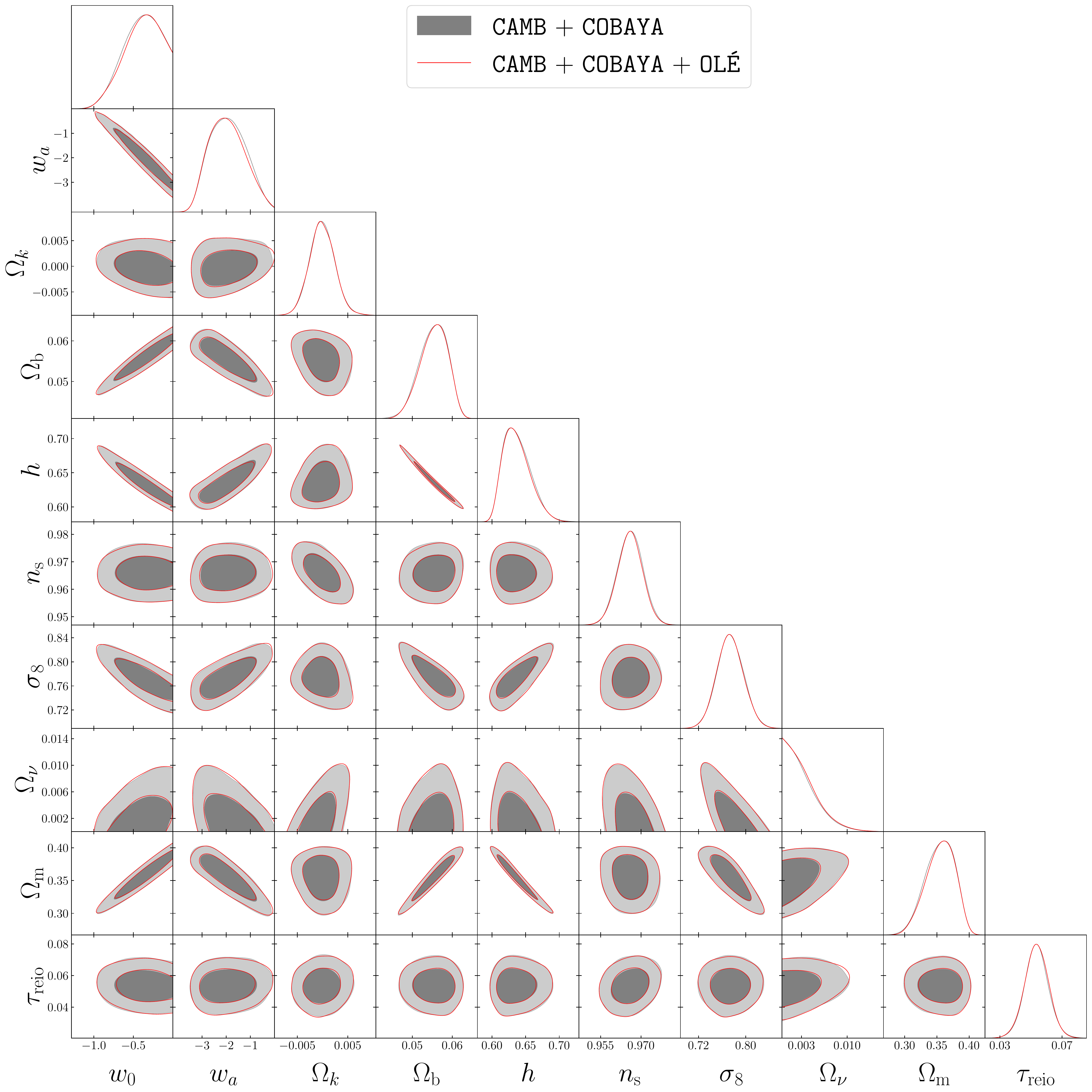}
    \caption{Posteriors of the extended cosmology model on \planck{} and 2024 DESI BAO data as outlined in example \ref{sec:extended}.
    Posterior means obtained using \ole{} and \camb{} differ by $D_x<0.048$, indicating good emulator accuracy.
    }
    \label{fig:extended}
\end{figure}

\vspace{0.1cm}

\noindent {\it Efficiency.} In this case, the runtime of \camb{} on 4 cores is $\sim 15$ seconds\footnote{This runtime is increased because we ask for high precision in the CMB spectra at high $\ell$, as required by the SPT likelihood. Note that using lower accuracy settings could result in at most 4 points shift in $\chi^2$ (see Appendix A.3 of~\cite{ACT_Precision}.},
while the evaluation of the likelihoods takes about $\sim 0.05$ seconds. Thus, Equation~\eqref{eq:speedup} suggests a maximum achievable speed-up of about $\sim 300$, again, neglecting the cost of building up the cache and testing the emulator accuracy in the \ole{} run. In practice, we observe that \ole{} increases the overall sampling rate from $0.5\, \mathrm{ESS/CPUh}$ to $173.1\, \mathrm{ESS/CPUh}$, boosting the entire MCMC run by a factor of 354. This is particularly impressive considering the fact that the \ole{} emulator does not require any pre-training.

\vspace{0.1cm}

\begin{figure}
    \centering
    \includegraphics[width=0.95\linewidth]{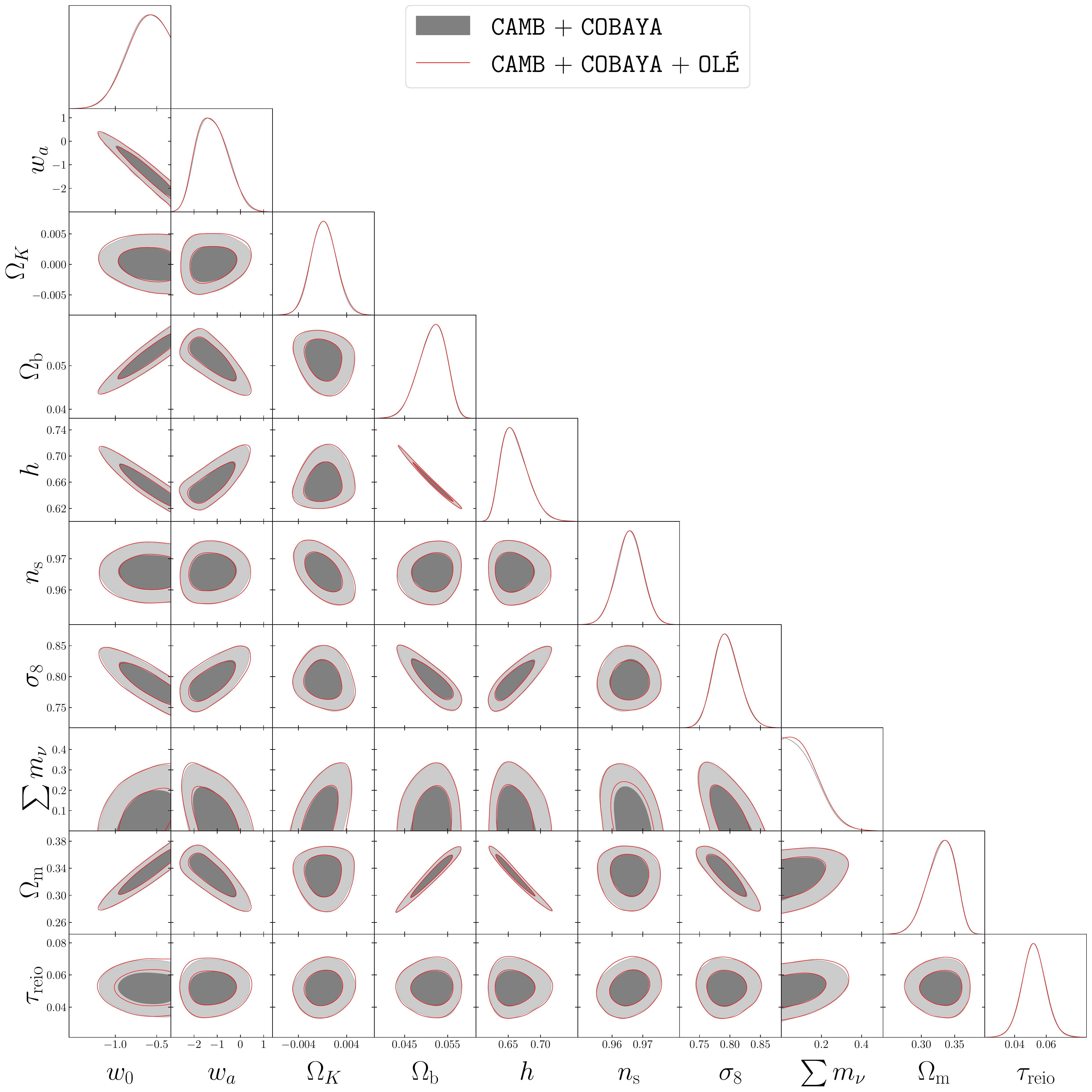}
    \caption{Posteriors of the extended cosmology model on \planck{}, SPT-3G, and 2024 DESI BAO data as outlined in example \ref{sec:extended}.
    Posterior means obtained using \ole{} and \camb{} differ by $D_x<0.02$, indicating excellent emulator accuracy.
    }
    \label{fig:extended_spt}
\end{figure}

\noindent {\it Adding nuisance parameters.} To test \ole's performance in even more extreme conditions, we stick to the same extended model, but we add CMB temperature and polarization data from SPT-3G 2018~\cite{SPT-3G:2022hvq} to that from \planck{} and 2024 DESI BAO. The SPT-3G 2018 likelihood includes several nuisance parameters, resulting in a final 64-dimensional parameter space (10 cosmological parameters, 21 \planck{} nuisance parameters, and 33 SPT-3G nuisance parameters). With such high dimensionality, MCMC runs become prohibitively expensive. Using \camb{}, the MH sampler of \cobaya{} and its default oversampling option, we find that by running 8 chains (on 4 cores each) we reach $|R-1|=0.02$ after 13 days and 11 hours. 
Instead, an equivalent run performed using \ole{} (with only 2 cores per chain and no oversampling) converges at the level of $|R-1| =0.02$ in just 39 hours.\footnote{We don't report the effective sampling rate in this case, because the CPU efficiency was not reported by the HPC cluster.}
The accuracy of the \ole{} run remains excellent, with $D_x<0.02$.

\vspace{0.1cm}

\noindent The examples of this section suggest that the accuracy and efficiency of \ole{} scale extremely well with the complexity of the physical model and with the number of nuisance parameters. Some runs that would have a considerable computational cost with a traditional MCMC approach become easily feasible using the on-the-fly \ole{} emulator, without compromising accuracy.

\subsection{Stage-IV LSS Forecast for an Extended Cosmology Using \montepython{} and \class{}}
\label{sec:euclid}

\noindent {\it Data.} As a further test of the emulator's applicability to different types of likelihoods and data, we run MCMC chains on synthetic data accounting for a Stage-IV galaxy survey, using a version of the mock Euclid likelihoods for \montepython{} described in References~\cite{Audren:2012wb,Sprenger:2018tdb,Euclid:2023pxu,Euclid:2024imf}. More precisely, like in Reference~\cite{Euclid:2024imf}, we use a set of likelihoods describing the power spectrum of weak lensing maps, photometric galaxy clustering maps, the cross-correlation between the two previous observables, and spectroscopic galaxy clustering maps (abbreviated as $\mathrm{WL}+\mathrm{GC}_\mathrm{ph}+\mathrm{XC}_\mathrm{ph}+\mathrm{GC}_\mathrm{sp}$).\footnote{We made some very minor changes to the existing public Euclid likelihoods, since unlike the default CMB and BAO likelihoods, these are not well-optimized for \ole{}'s automatic emulation. These changes are explained in Appendix~\ref{sec:MPinterface}, and the modified likehihoods will be included in a public \montepython{} release. Note that this does not change the fact that \ole{} works out of the box with both \cobaya{} and \montepython{}, requiring no modifications to existing likelihoods in nearly all cases.}
Both of the runs of this section are based on the MH sampler implemented in \montepython{}, with 12 parallel chains running for 65 hours. We run the standard \class{} chains with 8 cores per chain oversampling the nuisance parameters by a factor of five and the \ole{} chains on a single core per chain.
We perform the \ole{} run using precision parameters very similar to those of the previous subsection, with a looser-than-default requirement on the log-likelihood error $\Delta \ln {\cal L}$ below which the emulator is considered valid, and an enhanced accuracy setting for the number of PCA components (see Appendix~\ref{sec:precisions_euclid} for details).

\vspace{0.1cm}

\noindent {\it Model.} Like in the analysis of Reference~\cite{Euclid:2024imf}, we use an extended cosmology with four free parameters on top of the five standard $\Lambda$CDM parameters.\footnote{In absence of CMB data, we do not include the optical depth to reionization in the list of free $\Lambda$CDM parameters.} Our model includes the CPL parametrization of DE, see Equation~\eqref{eq:CPL}, massive neutrinos parameterized with the summed mass $\sum m_\nu$, as well as additional ultra-relativistic free-streaming relics parameterized through $\Delta N_\mathrm{eff}$. We adopt fiducial parameter values close to the {\it Planck} $\Lambda$CDM best-fit, with a pure cosmological constant, $\sum m_\nu=0.06$~eV, and no relativistic relics.

\vspace{0.1cm}

\noindent {\it Accuracy.} 
Visually, we find excellent agreement between the \class{}-only posteriors and those computed with the help of \ole{}, with normalized shifts in the means, $D_x\lesssim0.1$ for all parameters, with $D_x\sim0.01$ for the majority. A close inspection of Figure~\ref{fig:euclid_triangle} shows that in all our runs, there is a small shift between the mean posterior value and the fiducial value for $h$ and $n_s$. This is caused by the non-Gaussian nature of the posterior, as already found  in Reference~\cite{Euclid:2024imf} (see Fig.~13 therein).

\vspace{0.1cm}

\begin{figure}
    \centering
    \includegraphics[width=\linewidth]{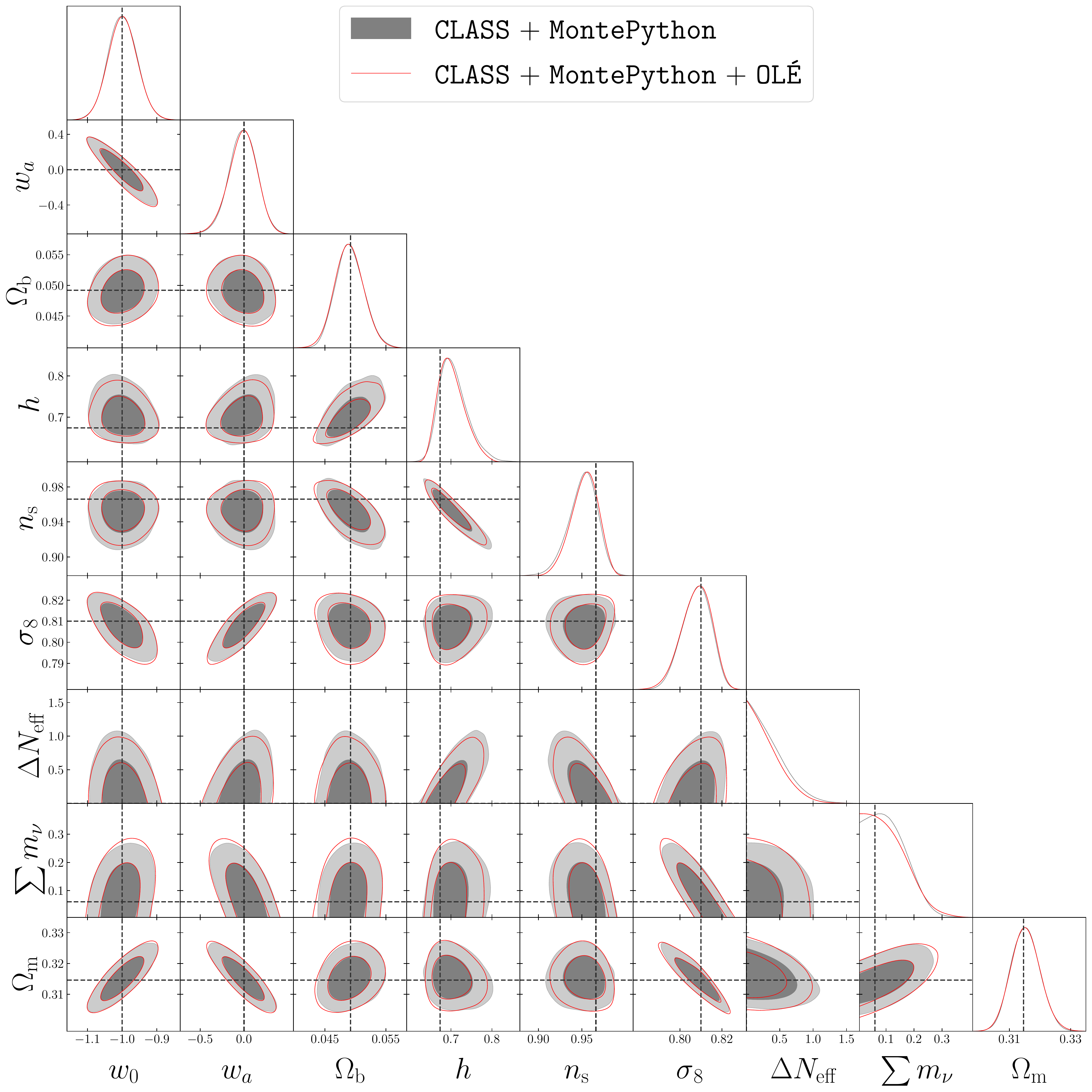}
    \caption{Expected posteriors of the $w_0 w_a \mathrm{CDM} + \sum m_\nu+N_\mathrm{eff}$ model according to the mock Euclid $\mathrm{WL}+\mathrm{GC}_\mathrm{ph}+\mathrm{XC}_\mathrm{ph}+\mathrm{GC}_\mathrm{sp}$ likelihoods described in Section~\ref{sec:euclid}, with a fiducial model close to the {\it Planck} $\Lambda$CDM best-fit. Dashed gray lines indicate the fiducial parameter values.
    We recover the same posterior means with or without the \ole{} emulator to within $D_x\lesssim 0.1$.
    Details on the precision parameters can be found in the Appendix~\ref{sec:precisions_euclid}.
    }
    \label{fig:euclid_triangle}
\end{figure}

\vspace{0.1cm}

\noindent {\it Efficiency.} In this case, we measure $T_{\cal T}\sim 19$~s and $T_{\cal L}\sim 0.32$~s on 4 cores. The maximum achievable speed-up according to Equation (\ref{eq:speedup}) is around $\sim$ 59, slightly smaller than in previous cases because the mock Euclid likelihoods are significantly slower than the {\it Planck}, supernovae and BAO likelihoods.\footnote{\class{} is also significantly slower in this case because the Euclid likelihoods require a calculation of the matter power spectrum up to a large wavenumber and over a fine grid. However, in the ratio, the effect of likelihoods being slower wins.} The ESS and the effective sampling rate ESS/CPUh found in runs with and without \ole{}  are presented in Table~\ref{tab:ESS_euclid}. They show that \ole{} speeds up the MCMC by a factor of $49$, close to the theoretical speed-up factor. Once more, we find that \ole{} has a decisive impact on the time and energy consumption of the runs.

\begin{table}[]
\centering
\begin{tabular}{lcc}
\hline
Run & ESS (mean) & ESS/CPUh \\ \hline
\class+\montepython      & 2204       & 0.35                  \\
same+\ole{}  & 13541      & 17                    \\ \hline
\end{tabular}
\caption{Mean effective sample size (ESS) across MCMC chains and effective sampling rate (ESS/CPUh) for our mock Euclid forecasts, in a traditional run and in an \ole{} run. 
For details on the \ole{} precision parameters see Appendix~\ref{sec:precisions_euclid}.}
\label{tab:ESS_euclid}
\end{table}

\subsection{Early Dark Energy with Current Data Using \montepython{} and \texttt{TriggerCLASS}}
\label{sec:NEDE}

\noindent {\it Model.} As an example of a cosmology with strongly non-Gaussian posteriors, we use a model of New Early Dark Energy (NEDE), which has been studied recently in the context of cosmological tensions \cite{Chatrchyan:2024xjj,Garny:2024ums}. This analysis relies on the theory code \texttt{TriggerCLASS}.\footnote{\url{https://github.com/NEDE-Cosmo/TriggerCLASS}} In addition to the standard $\Lambda$CDM parameters, the NEDE model introduces four extra free parameters, \{$f_\mathrm{NEDE}$,  $\Omega_\phi$, $\log_{10}z_*$, $w_*$\}, accounting respectively for the fraction of NEDE at decay time, the density of the trigger field, the redshift at decay time, and the equation of state after the decay. We fix the summed neutrino mass to $\sum m_\nu=0.06$~eV.

\vspace{0.1cm}

\noindent {\it Data.} Our analysis includes \planck{} 2018 data for temperature, polarization~\cite{Planck:2019nip} and lensing~\cite{Planck:2018lbu} anisotropies, as well as Supernova luminosity data from the Pantheon+ sample~\cite{Brout:2022vxf} and BAO data from the SDSS DR7~\cite{Ross:2014qpa} and DR12~\cite{BOSS:2016wmc} samples. We use the MH sampler of \montepython{}. We run 16 chains of 400,000 points each, on 4 cores each in the traditional run and 1 core each in the \ole{} run. 

\begin{figure}[ht]
    \centering
    \includegraphics[width=0.9\linewidth]{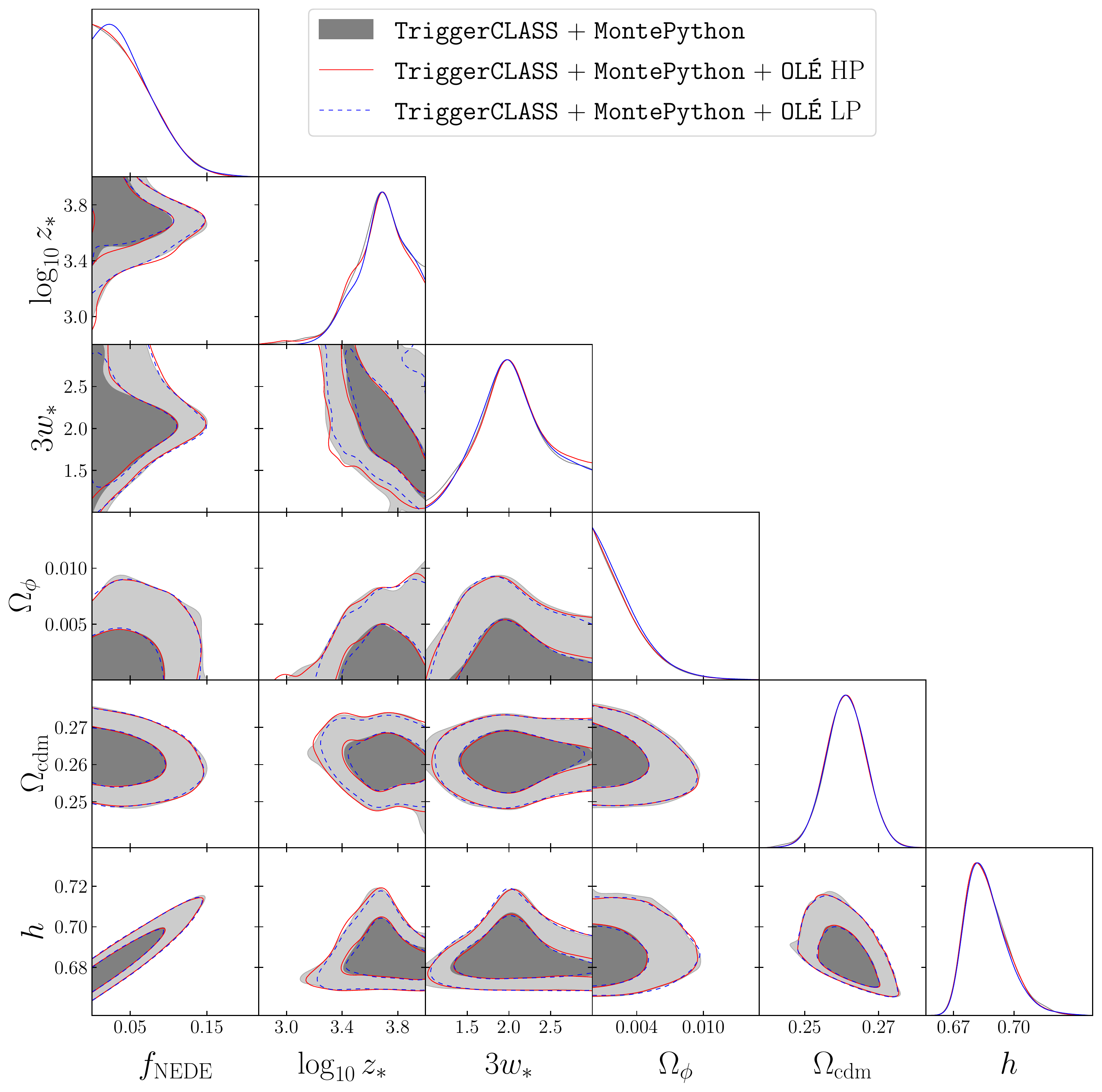}
    \caption{Posteriors of  NEDE parameters along with $\Omega_\mathrm{cdm}$ and $h$. Despite the strongly non-Gaussian shapes of the posteriors, \ole{} does an excellent job of recovering the results of the standard MCMC analysis.
    }
    \label{fig:NEDE}
\end{figure}

\vspace{0.1cm}

\noindent {\it Accuracy.} On top of experimenting the NEDE model, we use this section to test the impact of a few precision settings in \ole{}. In appendix \ref{sec:precisions_NEDE}, we define two sets of \ole{} accuracy parameters that we call high precision (HP) and low precision (LP). These parameters control the criterion for considering the emulator as accurate enough in a given point and the number of PCA components. The LP settings degrade the default requirements on the emulator accuracy and on the number of PCA components. In the HP settings, the requirements on the emulator accuracy are stronger than in the default settings, but the number of PCA components is in between the default and LP settings. With both settings, we find excellent agreement between the posteriors obtained with and without \ole, as illustrated in Figure~\ref{fig:NEDE}. We do not compute the shifts in the means $D_x$ for this model, since normalizing to the standard deviation is less meaningful when the posterior shapes are so far from Gaussian. Instead we wish to stress the very high level of agreement of the posteriors, which is evident on visual inspection of the contours. This shows that the \ole{} emulator is able to train over a non-trivially-shaped region of the parameter space and to accurately capture all the subtle physical effects of the NEDE model. Additionally, the stability of the results against different accuracy settings (HP and LP) shows that our error auto-evaluation scheme is robust and that \ole{} does not require a fine-tuning of its precision parameters in order to remain accurate.

\begin{table}[ht]
    \centering
    \begin{tabular}{lcc}
        \hline
        Run & ESS (mean) & ESS/CPUh (norm.) \\
        \hline
        \class+\montepython   & 24949.04       &  1.2 \\
        same+\ole{} (LP)   & 66670.37     &  31.2  \\
        \hline
    \end{tabular}
 \caption{Mean effective sample size (ESS) across MCMC chains and effective sampling rate for our NEDE analysis, in a traditional run and in an \ole{} run using LP. For details on the \ole{} precision parameters see Appendix~\ref{sec:precisions_NEDE}.
 }
    \label{tab:NEDE_ESS}
\end{table}

\vspace{0.1cm}

\noindent {\it Performance.} In this example, we find $T_{\cal T}\sim 4$~s\footnote{In Section \ref{sec:lcdm} we quoted a runtime of 5~s for \class. The small difference just comes from using slightly faster processors in the runs of this section.} and $T_{\cal L}\sim 0.08$~s on 4 cores. The running time of the likelihood is increased compared to Section \ref{sec:lcdm} because the Pantheon+ likelihood is slower than the older Pantheon one. The maximum achievable speed-up according to Equation (\ref{eq:speedup}) is 50, smaller than in Section \ref{sec:lcdm} due to the slower likelihoods.
The \ole{} run with LP is accelerated by a factor $\sim$31. Even in this extreme case, \ole{} is able to save between one and two orders of magnitude in runtime and energy consumption while providing accurate results.

\section{Staged NUTS Sampling with Differentiable Likelihoods}
\label{sec:diff_like}
In this Section, we interface \ole{} with the CMB likelihood library \candl{}.\footnote{\url{https://github.com/Lbalkenhol/candl}}
While \ole{}'s emulators calculate CMB spectra from a set of cosmological parameters, \candl{} modifies the CMB spectra to account for foregrounds, calibration, and other systematic effects, and calculates the likelihood value for a chosen data set.
Crucially, both codes are written with \texttt{JAX} support, which features an automatic differentiation algorithm; \ole{} and \candl{} together form a fully differentiable pipeline with quick and easy access to derivatives of the log-likelihood value with respect to cosmological and nuisance parameters.
This allows for the use of more efficient minimization and MCMC sampling algorithms \citep[see e.g.][]{nash84, duane87, nocedal06, Hoffman:2011ukg, kingma14, deepmind20, robnik22}.

We develop the following inference strategy to exploit the differentiability of \ole{} models and \candl{} likelihoods:
\begin{enumerate}
    \item Run MH sampling using a full Boltzmann solver with multiple chains in parallel, saving the predictions to the cache.
    \item Once sufficient points have been collected, train and switch over to the emulator.
    \item Minimize the posterior to find the best-fit point using the gradient-based truncated Newton algorithm implemented in \texttt{Scipy}. \citep{nash84, nocedal06, scipy}
    \item Perform gradient-based No U-Turns (NUTS) \citep{Hoffman:2011ukg} sampling of the posterior.
\end{enumerate}
We refer to this prescription as \textit{staged NUTS sampling} and visualize the procedure in Figure~\ref{fig:staged_sampler}.
Note that during NUTS sampling we can still fall back to occasional evaluations of the Boltzmann solver and re-training of the emulator if its accuracy is insufficient.
In fact, the testing of the emulator accuracy is improved when using a differentiable likelihood as the emulator uncertainty can simply be propagated through to the posterior value, without having to draw multiple realizations of the observables and evaluating the corresponding posterior values individually.

To assess the performance of staged NUTS sampling compared to regular MH sampling with \ole{} we compare the following two cases:
(1) MH sampling with eight parallel chains with eight CPUs each using \cobaya{}, no oversampling of nuisance parameters and
(2) staged NUTS sampling as described above across eight CPUs.
In both cases we explore \LCDM{}, using \class{} as the Boltzmann solver and the differentiable SPT-3G 2018 \TTTEEE{} lite likelihood \citep{Balkenhol:2024obe, SPT-3G:2022hvq} available in \candl{} \citep{Balkenhol:2024sbv}.
Both chains start with the same sub-optimal proposal matrix.

\begin{figure*}
    \centering
    \includegraphics[width=1.0\columnwidth]{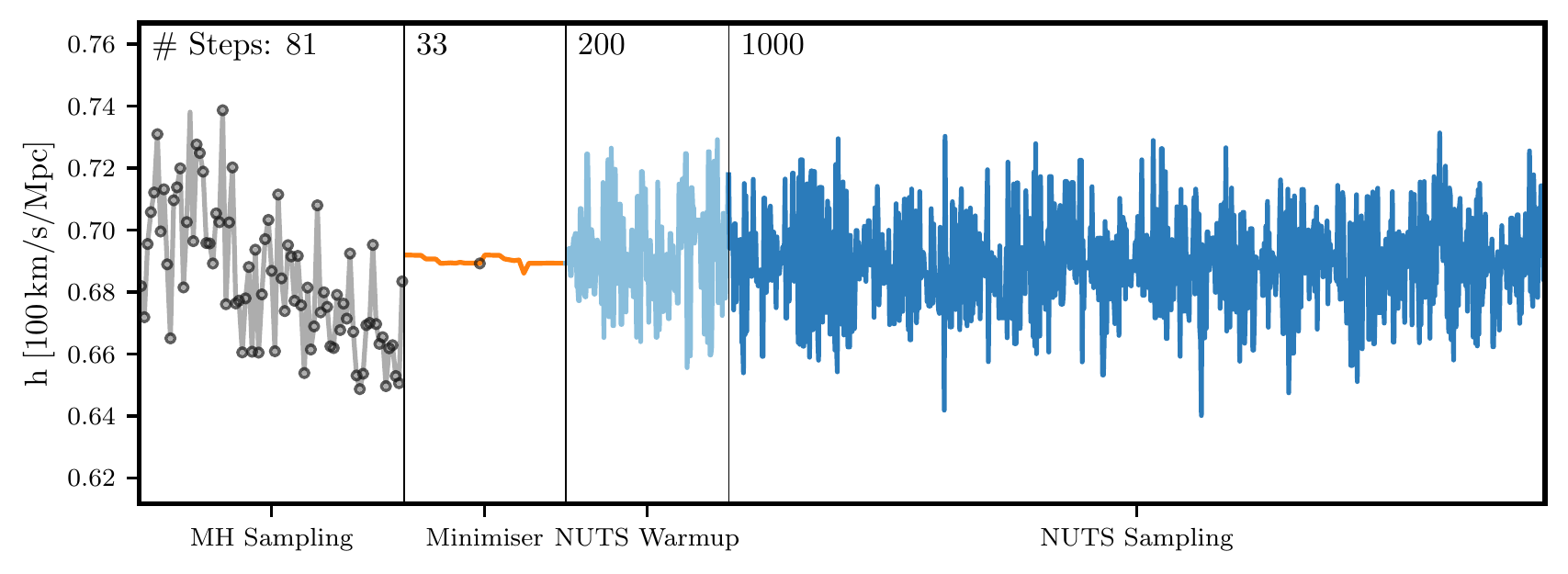}
    \caption{Example chain using the staged NUTS sampler.
    We explore \LCDM{} using the SPT-3G 2018 \TTTEEE{} lite likelihood and show $h$ samples. 
    We indicate the phases of the staged sampler along the horizontal axis; from left to right: MH sampling (gray), minimization (orange), NUTS warmup sampling (light blue), NUTS sampling (blue).
    Points retained for emulator training have been highlighted with open black circles.
    Note that the x-axis scale is arbitrary and has been chosen for visualization purposes;
    we indicate the number of sample points in each stage along the top.
    This inference strategy exploits the differentiability of \ole{} and \candl{};
    burn-in is effectively complete after the minimization step and the subsequent NUTS sampling explores the posterior more efficiently than traditional MH sampling.}
    \label{fig:staged_sampler}
\end{figure*}

We find that NUTS sampling explores the posterior distribution more efficiently: the ESS/CPUh after the burn-in phase is a factor of $4.0$ higher for our staged NUTS sampler than for the traditional MH chains.
This indicates more informative samples are generated at a faster rate and intuitively means that staged NUTS chains with the same information content as MH chains can be run in one quarter of the time, provided the same hardware and energy allocation.
This is similar to what was found when comparing NUTS and MH sampling by \citep{Balkenhol:2024sbv} for CMB data from the Atacama Cosmology Telescope \citep{ACT:2020gnv, ACT:2020frw} and by \citep{Mootoovaloo2024} for tomographic $3\times2$ point analyses.
However, we stress that in these cases the emulators used were trained ahead of time using significant computational resources, whereas here we generate them on the fly with comparatively little cost.

Furthermore, we note that the time it takes to arrive at posterior-representative sampling, i.e. the burn-in, is efficiently used and elegantly defined for our staged NUTS sampler.
Burn-in is completed with step (3) in the description above and since minimization is fast when gradient information is available, the burn-in time is effectively confined to how long it takes to gather training points and deploy the emulator.
When this can be done depends on the complexity of the model being explored, though we recommend $80$ points as a default option.
This is lower than what is typical for MH-based samplers, though we note that in this case the burn-in definition varies depending on the specific implementation.
For \cobaya{}, it is usually recommended to throw away the initial $20-30\%$ of points, which scales unfavorably with the runtime due to the proportionality with the chain length.\footnote{See \url{https://cobaya.readthedocs.io/en/stable/sampler_mcmc.html}}
For \montepython{}, burn-in is regarded as complete once chains have reached the vicinity of the best-fit point, which is sensitive to the quality of the starting point and the initial proposal matrix as well as the shape of the posterior distribution.

\section{Conclusion}
\label{sec:conclusion}
In this work, we have presented the emulator code \ole{}, including its design goals and architecture. We have demonstrated its performance and accuracy, achieving comparable results to traditional Einstein-Boltzmann solvers \class{} and \camb{} at a fraction of the computational cost across a range of cosmological models and datasets. \ole{} relies on active sampling and online learning to efficiently acquire training points with high information content on the fly, requiring no pre-training and outperforming similar emulators in data efficiency. It is released with its own built-in sampler as well as user-friendly interfaces for both the \montepython{} and \cobaya{} samplers, allowing for easy integration in existing inference pipelines.
We showcased the performance on a variety of cosmological data sets including measurements from \planck{}, SPT-3G, Euclid mock data, BAO and supernovae data. The tested cosmological models range from the 6 parameter $\Lambda$CDM model over different extensions to a New Early Dark Energy model with 10 parameters with highly non-Gaussian posterior distributions. We find a computational speed-up in the parameter inference process ranging from one to two orders of magnitude, depending on the considered model and data set. It is close to the maximum achievable speed-up expected when considering the evaluation of \camb{} or \class{} as the bottleneck of the inference process. Lastly, we use \ole{} together with the CMB likelihood library \candl{}. We exploit the differentiability of both codes to construct a staged NUTS sampler that uses the gradient of the likelihood to explore the posterior distribution approximately four times more efficiently than traditional MH sampling.
\ole{} is publicly available and we release a series of examples that show how to use the code.\footnote{\url{https://github.com/svenguenther/OLE}}

The application of machine-learning techniques is becoming wide-spread in cosmology and allows us, as demonstrated here, to accelerate common computational tasks.
This facilitates the broadening and deepening of cosmological analyses by exploring constraints placed by many variations of the data across a broad range of models.
\ole{} thus represents an advancement in numerical methods demanded by the precision of contemporary cosmological data.
As the evaluation time of the theory code becomes subdominant, it is pertinent to improve the efficiency of our samplers and likelihood codes in the future.
This development is particularly timely given the environmental impact of high-performance computing.

\appendix
\section{Using OLÉ}
\label{sec:usingOLE}
The \ole{} code has been designed to be easy to use, either as a full standalone pipeline, or as a supplement to accelerate the theory code in existing pipelines. For this purpose, \ole{} includes its own samplers, as well as interfaces for the the \cobaya{} and \montepython{} samplers.

As \ole{} automatically copies the functionality of the theory code, it should work out of the box with any modified version of \class{} or \camb{} that do not change the structure of calls to the wrapper required to compute the observables. For example, we make use of \texttt{TriggerCLASS} in Section~\ref{sec:NEDE}, which is one such modified version of \class{}.

\subsection{Implemented Sampler}

In addition to the emulator, \ole{} includes a set of integrated sampling methods. Alongside an ensemble sampler based on the Python package \texttt{emcee} \cite{Foreman-Mackey:2012any}, it features a minimizer utilizing \texttt{Scipy} \cite{Virtanen:2019joe} and a gradient-based NUTS sampler. An example demonstrating the combination of these methods is provided in Section \ref{sec:diff_like}.

The implementation of theory codes is inspired by \cobaya{}. A theory code is implemented as a \textit{child class} of the \ole{}-internal \texttt{Theory} class. It has an attribute \texttt{requirements}, which represents the set of quantities required by the likelihood. This attribute is dynamically determined by all likelihoods within the sampler. 

The theory class also provides the function \texttt{compute}, which takes a dictionary of parameters and populates it with the required quantities. The likelihoods are implemented as \textit{child classes} of the \texttt{Likelihood} class. Each likelihood child class must define the \texttt{loglike} function, which takes the dictionary of computed quantities and evaluates the log-likelihood. Additionally, it includes the function \texttt{update\_theory\_settings}, which allows enforcing specific settings onto the theory code. If the \texttt{loglike} function is implemented using \texttt{JAX} and is differentiable, gradient-based methods such as NUTS or a gradient-based minimizer can be employed.

Currently, we provide interfaces for \class{} and \camb{}, as well as likelihood implementations for \candl{} and BAOs.\footnote{Examples can be found at \url{https://github.com/svenguenther/OLE/tree/main/OLE/examples/candl}.}

\subsection{\montepython{} Interface}
\label{sec:MPinterface}
The \montepython{} interface allows \ole{} to (partially) replace \class{} when using \montepython{}, relying on the \ole{} emulator while letting \montepython{} handle the sampling and likelihood calls. The interface allows \ole{} to work out of the box with existing likelihoods. It only requires setting the path to your existing \montepython{} installation in the MP\_PATH configuration file inside your \ole{} installation, and executing the interface as you would normally execute \montepython{},
\begin{minted}[
    gobble=4,
    frame=single
  ]{shell}
    python /path/to/OLE/interfaces/montepython_interface.py -p your_MCMC.param 
\end{minted}
including any additional \montepython{} arguments at the end. No changes to existing input parameter files are needed, but the emulator settings to pass to \ole{} can be set in the input parameter file via the new dictionary \texttt{data.emulator\_settings}.

The interface operates by inserting \ole{} as an intermediate layer between \montepython{} and the cosmology code, in this case usually \class{}. It creates copies of all functions in the Python wrapper \texttt{classy} and constructs emulators for those functions utilized in the likelihoods specified in the input parameter file. These emulators then approximate the output of these functions and provide the results to each likelihood code. 
To ensure accurate emulation, the interface classifies the \texttt{classy} wrapper functions based on their input and output types, allowing the emulator to correctly reproduce their behavior.
The most commonly used functions suitable for \ole{} are listed in \texttt{montepython\_interface.py}. {\color{blue} \sout{However, i}}If any new functions suitable for emulation are added {\color{blue} to the \texttt{classy} wrapper}, they should be added to this list{\color{blue}, however, most \class{} modifications do not require the addition of new functions to the wrapper, in which case this step is not necessary}.

Even if the interface allows \ole{} to work out of the box with existing likelihoods, it is worth considering whether a given likelihood is written in the optimal way in regards to emulation.
Many existing \montepython{} likelihoods utilize functions that return both the cosmological results on the grid used internally by \class{} and the grid itself. For example, there are function returning both $P_\mathrm{m}(k,z)$ and a $(k,z)$-grid, or the CMB $C_\ell$s along with a one-dimensional $\ell$-grid. In such cases, \ole{} would, by default, emulate both the desired quantity and the coordinate values of the grid points.
In addition to the computational overhead of emulating these outputs, this introduces extra uncertainty from the emulator, and a priori \ole{} does not have any protection against non-nonsensical coordinate values, e.g. multiple identical $k$-values or non-integer $\ell$s. 

In the case of $C_\ell$s, this can be solved trivially by passing the names of outputs we are not interested in via the \texttt{`skip\_emulation\_quantities'} parameter.

In the case of large scale structure functions, the easiest solution is to replace in the likelihood codes any call to a function returning $k$ and $z$ grids in addition to power spectra by equivalent functions returning $P(k,z)$ on an input grid, which can be defined in the likelihood's \texttt{.data} file. This requires only minor modifications to existing likelihoods, and the equivalent functions are already defined in \texttt{classy}. We provide examples of this in the form of the \texttt{euclid\_photometric\_alm\_OLE} and \texttt{euclid\_spectroscopic\_OLE} likelihoods used in Section~\ref{sec:euclid}, which are equivalent to the existing public \montepython{} likelihoods without the \texttt{\_OLE} suffix as used in Reference~\cite{Euclid:2024imf}.  Using these likelihoods as an example, the concrete changes that were made to the likelihoods and the \texttt{classy} wrapper are as follows:
\begin{itemize}
    \item We replaced calls to \texttt{classy} functions of the type \texttt{get\_pk\_and\_k\_and\_z()} with calls of the type \texttt{get\_pk()}. In practice, this means the interpolation in the $P(k,z)$-grid computed by \class{} happens in the wrapper, rather than in the likelihood.
    \item The existing public likelihoods get the reduced Hubble parameter from a call to the \texttt{classy} function \texttt{h()}. Thus, \ole{} may automatically try to emulate this function. Instead, we implement a check to see if this parameter is in the list of MCMC parameters, and can thus be obtained directly from the sampler. The function \texttt{h()} is used and emulated by \ole{} only when this is not the case. Such checks can be implemented for other background cosmological parameters needed in a given likelihood.
    \item The existing likelihoods call two \texttt{classy} functions to get the total matter spectrum $P_\mathrm{m}(k,z)$ and the baryon plus CDM power spectrum $P_\mathrm{cb}(k,z)$, before taking their ratio. It is more efficient for \ole{} to emulate directly the ratio instead of the numerator and the denominator. For this purpose, we added to \texttt{classy} a function \texttt{get\_Pk\_cb\_m\_ratio()} returning directly the ratio. For this specific case, an additional function, \texttt{get\_tk()}, was also implemented, which returns a given transfer function on a provided $(k,z)$-grid, following the conventions of the \texttt{get\_pk()}-type functions.
\end{itemize}

Finally, as explained in Section~\ref{sec:OLE_mcmc}, after the initial emulator training phase, \ole{} automatically disables oversampling and the fast/slow parameter decomposition by setting the \montepython{} argument \texttt{--jumping} to \texttt{global} instead of the default \texttt{fast}.

\subsection{Cobaya Interface}
The \cobaya{} interface of \ole{} allows the use of the \ole{} algorithm inside the \cobaya{} framework with all included likelihoods, theory codes and samplers. The implementation is such that the \cobaya{} code files and the overall usage of \cobaya{} remain unchanged. \ole{} only modifies parts of the code once they are loaded in memory at the time of execution.

The usage of the emulator can be turned on by giving the \texttt{theory}-element in the initial \texttt{yaml}/\texttt{dict} the keyword \texttt{emulate} with the value \texttt{true}. The keyword \texttt{emulator\_settings} takes a dictionary of settings to change the default \ole{} settings.\footnote{Examples are available at \url{https://github.com/svenguenther/OLE/tree/main/OLE/examples/Cobaya}}
When implementing an inference pipeline with the \cobaya{} internal \texttt{Theory} classes, it is sufficient to import them from \texttt{OLE.interfaces.cobaya\_interface} while all other components can remain as plain \cobaya{}.\footnote{For detailed documentation on building \texttt{Theory} and \texttt{Likelihood} classes, see \href{https://cobaya.readthedocs.io/en/latest/}{\texttt{https://cobaya.readthedocs.io/en/latest/}}.}
However, there remain a few points to consider when using the \cobaya{}-\ole{} interface. These are:

\begin{itemize}
    \item The interface works when \cobaya{} calls one theory code only (e.g., either \camb{} or \class). The user can involve multiple theory codes in their pipeline (e.g. a Boltzmann code and a Nucleosythesis code) only if they are nested into each other (e.g., \texttt{Primat} \cite{Pitrou:2019nub} could still be called from inside \class).
    \item As outlined in section \ref{sec:OLE_mcmc} oversampling is useful in the early \textit{burn-in} stage to converge to the relevant samples more quickly. However, once the emulator is trained, the evaluation time of the emulator becomes typically shorter than that of the likelihood. Then, it is useful to switch oversampling off. This can be done by providing the MCMC sampler with the options \texttt{'callback\_function': OLE\_callback\_function} and \texttt{'callback\_every': 1}.
    \item Due to the particular implementation of \camb{},\footnote{The theory code of \camb{} is split between \texttt{camb} and \texttt{camb\_transfers}. This allows oversampling for some of the cosmological parameters. However, this behavior is not supported by \ole{} and can lead to errors.} it is required to manually set a blocking of the cosmological and nuisance parameters, as in the example script shown below. Further examples are available on the \ole{} GitHub page.
    \item The interface has been tested for a variety of likelihoods, including CMB, BAO, lensing and SNIa likelihoods. When using likelihoods with an unusual input/output format, we strongly recommend to validate the results first before using \ole{}. Note that emulating observables with varying size is not implemented in \ole{}.
\end{itemize}

\noindent Eventually, the script to start an inference run barely changes:

\begin{minted}[
    gobble=4,
    frame=single,
    linenos
  ]{python}
    # import OLE-cobaya interface for your Cobaya version
    from OLE.interfaces.cobaya_interface import *

    info = {
        'theory': {
          'classy': {
            'emulate': true # necessary line to request emulation
            'emulator_settings': {...}, # optional change of default settings
              ...},
        'likelihood': {
          'planck_2018_highl_plik.TTTEEE': {},
          ... },
        'params': {
          ... },
        'sampler': {
          'mcmc': {
            'callback_function': OLE_callback_function, # see 'callback_every'
            'callback_every': 1, # to disable oversampling once trained
            'blocking': [1,['H0',...,'ns']], [5,['A_planck',...]]], # CAMB only
          }, }, }
          
    # run MCMC
    updated_info, sampler = cobaya.run(info)
    \end{minted}

\section{Precision Settings}
\label{sec:precsions}

The accuracy of the \ole{} emulator is determined by a set of precision parameters that can either be specified by the user or take default values. For most of the examples presented in Section~\ref{sec:examples}, we have conducted internal tests with multiple sets of \ole{} precision settings to examine the relationship between these settings, accuracy, and computational cost. 
For completeness, this appendix lists the precision parameters that were set to non-default values in the presented examples.

\subsection{Default Precision Parameters}
\label{sec:precisions_default}

We present an overview of the most important precision parameters, followed by their default value in bold. For a complete list, please refer to the online documentation.

\begin{itemize}
    \item \oleparam{cache\_size}{500}\\
    Maximum number of stored training data points. If more data points are to be added, the one with the smallest log-likelihood is removed.
    \item \oleparam{min\_data\_points}{80}\\
    Minimum number of points in the cache before the emulator can be trained. If this number is too small, the emulator will require many re-trainings. If it is too large, the initial data-gathering phase of \ole{} is unnecessarily long.
    \item  \oleparam{delta\_loglike}{50}\\
    This parameter sets the threshold between relevant data points (to be cached) and outliers. All points in the cache whose log-likelihood differs from the maximum log-likelihood by more than \texttt{delta\_loglike} are classified as outliers and removed. If \texttt{N\_sigma} and \texttt{dimensionality} (see below) are set, this parameter is ignored.
    \item \oleparam{dimensionality}{None}\\
    As an alternative to \texttt{delta\_loglike}, we can infer the threshold between relevant points and outliers by computing the $\Delta\log\mathcal{L}$ of a Gaussian distribution (with a dimensionality set by the parameter \texttt{dimensionality}) from its best fit point to \texttt{N\_sigma} standard deviations. Thus, if the posterior is approximated by a Gaussian, all points in the cache lay inside the \texttt{N\_sigma} contour, while points outside are classified as outliers. This parameter should be set to the total number of MCMC parameters (cosmology and nuisance parameters). If no dimensionality is specified, \texttt{delta\_loglike} is used, but in general it is recommended to pass \texttt{N\_sigma} and \texttt{dimensionality}.
    \item \oleparam{N\_sigma}{3.0}\\
    This parameter sets the range around the best-fit point within which points should be cached rather than classified as outliers. To be used in conjunction with \texttt{dimensionality}.
    \item \oleparam{min\_variance\_per\_bin}{$\mathbf{5\times10^{-6}}$}\\
    The level of compression of each observable depends on the number of PCA components. To determine this number, \ole{} increases the number of PCA components until the explained variance per bin times the bin size exceeds this parameter. Setting this variance to $\sigma^2 = 10^{-4}$ would mean that for each observable, the truncation of the expansion into PCA components introduces a typical relative error on the normalized observable of $\sigma \sim 10^{-2}$. Thus, this parameter accounts for the maximal achievable precision of the emulator. If it is too large, results might be biases. In presence of highly correlated parameter effects, it is advisable to reduce this number by one or two orders of magnitude.
    \item \oleparam{quality\_threshold\_constant}{0.1}\\
    In addition to the compression of the PCA, the second major source of error in \ole{} is the precision of the GP emulation. This is controlled by the following three precision parameters. First, \texttt{quality\_threshold\_constant} is the
    constant term for determining the maximum allowed error on the log-likelihood, as in Equation~\ref{eq:prec}.
    \item \oleparam{quality\_threshold\_linear}{0.05}\\
    This is the linear term for determining the maximum allowed error on the log-likelihood in Equation~\ref{eq:prec}.
    \item \oleparam{quality\_threshold\_quadratic}{$\mathbf{10^{-4}}$}\\
    This is the quadratic term for determining the maximum allowed error on the log-likelihood in Equation~\ref{eq:prec}. When using \texttt{dimensionality} and \texttt{N\_sigma}, this parameter is ignored and the quadratic term is computed automatically in such way that points beyond \texttt{N\_sigma} are considered as outliers.
    \item \oleparam{N\_quality\_samples}{5}\\
    The number of samples drawn from the emulator to estimate its own accuracy. Large values result in repeated likelihood evaluations and slow the code down. Small values increase the degree of randomness in the evaluation of the emulator accuracy, which may lead to adding unnecessary points to the cache.
    
\end{itemize}

\subsection{Extended Cosmology Example}
\label{sec:precisions_camb}

In Section \ref{sec:extended}, in order to get faster convergence and a smaller data set, we relax the precision parameter \texttt{quality\_threshold\_linear} to $0.1$. For parameters such as $\Omega_\nu$, whose posterior intersects the prior boundary $\Omega_\nu>0$ and whose effect on CMB observables is far from linear near $\Omega_\nu=0$, we want to have a very precise reconstruction of the observable. Thus, we set the sensitivity for the PCA compression to $\texttt{min\_variance\_per\_bin}$ to $10^{-6}$.

\subsection{Stage-IV LSS Forecast Example}
\label{sec:precisions_euclid}
For the example in \S \ref{sec:euclid} we use once more \oleparam{min\_variance\_per\_bin}{$10^{-6}$} and relax the precision parameter  \texttt{quality\_threshold\_constant} to ${1.0}$ for faster convergence.

\subsection{Early Dark Energy Example}
\label{sec:precisions_NEDE}

\begin{table}[H]
    \centering
    \begin{tabular}{|l|c|c|}
        \hline
        & \textbf{High Precision} & \textbf{Low Precision} \\
         \hline
        \texttt{quality\_threshold\_constant} & 0.1 & 0.4 \\
        \texttt{quality\_threshold\_linear}  & 0.01 & 0.1 \\
        \texttt{min\_variance\_per\_bin} & $10^{-5}$ & $10^{-4}$ \\
        \hline
    \end{tabular}
    \caption{\ole{} precision settings different from default values for the NEDE example of Section~\ref{sec:NEDE}.}
    \label{tab:settings_NEDE}
\end{table}

\subsection{Staged NUTS Sampling Example}
\label{sec:precisions_NUTS}

For the example in \S \ref{sec:diff_like} we set \oleparam{min\_variance\_per\_bin}{$\mathbf{10^{-6}}$} and \oleparam{N\_sigma}{4.0}.

\section{Cosmological Parameters}
\label{sec:par_def}

Table~\ref{tab:par_def} summarizes all cosmological parameters referenced throughout this work.

\begin{table}[H]
    \centering
    \begin{tabular}{|c|l|}
    \hline
    \textbf{Parameter} & \multicolumn{1}{|c|}{\textbf{Definition}}\\
    \hline
        $h$ & Expansion rate today $H_0$ in units of $100\,\mathrm{km/s/Mpc}$\\
        $\Omega_{\mathrm{b}}$ & Baryon fractional density\\
        $\Omega_{\mathrm{cdm}}$ & Cold dark matter fractional density\\
        $\ln(10^{10} A_{\mathrm{s}})$ & Amplitude of the power spectrum of initial scalar fluctuations\\
        $n_{\mathrm{s}}$ & Tilt of the power spectrum of initial scalar fluctuations\\
        $\tau_\mathrm{reio}$ & Optical depth to reionization\\
        $w_0, w_a$ & Dark Energy equation of state parameters in Equation~\eqref{eq:CPL}\\
        $\Omega_k$ & Mean spatial curvature parameter\\
        $\sigma_8$ & R.m.s. of matter fluctuations in a sphere of comoving radius of $8\ h^{-1}$ Mpc\\
        $\Omega_\nu$ & Neutrino fractional density\\
        $\Omega_\mathrm{m}$ & Total matter fractional density\\
        $\Delta N_{\mathrm{eff}}$ & Deviation of the number of effective neutrino species from the \LCDM{} prediction\\
        $\sum m_\nu$ & Sum of neutrino masses\\
        $f_{\mathrm{NEDE}}$ & Fraction of NEDE at the time of decay\\
        $\log_{10}{z_\ast}$ & Logarithm of the redshift at NEDE decay time\\
        $w_\ast$ & Equation of state of the NEDE fluid after decay\\
        $\Omega_\phi$ & Density of the NEDE trigger field\\
         \hline
    \end{tabular}
    \caption{Brief definitions of all cosmological parameters that appear in the manuscript.}
    \label{tab:par_def}
\end{table}

\section{Introduction to Gaussian Processes}
\label{app:GP}

Gaussian Processes (GPs) are a non-parametric framework for modeling distributions over functions. They are especially well-suited for regression, classification, and optimization tasks where quantifying uncertainty is important.

\subsection{Overview}

A Gaussian Process is a collection of random variables, any finite number of which have a joint Gaussian distribution. Formally, a GP is defined as:

\[
f(x) \sim \mathcal{GP}(m(x), k(x, x'))
\]

where:
\begin{itemize}
    \item \( m(x) = \mathbb{E}[f(x)] \) is the \textit{mean function}
    \item \( k(x, x') = \mathbb{E}[(f(x) - m(x))(f(x') - m(x'))] \) is the \textit{covariance (kernel) function}
\end{itemize}

In practice, the mean function is often assumed to be zero, \( m(x) = 0 \), without loss of generality. The choice of kernel \( k(x, x') \) determines the smoothness and generalization properties of the model.

\subsection{Gaussian Process Regression}

Given data \( \mathcal{D} = \{(x_i, y_i)\}_{i=1}^n \) with observations \( y_i = f(x_i) + \epsilon_i \), where \( \epsilon_i \sim \mathcal{N}(0, \sigma_n^2) \), Gaussian Process regression involves placing a GP prior on \( f(x) \) and computing the posterior distribution over functions. We write the set of arguments $x_i$ of the data set $\mathcal{D}$ as $X$. The set of points at which we want to predict the function $f$ is denoted as $X_*$.

The joint prior over the training outputs \( \mathbf{f} \) and test outputs \( \mathbf{f}_* \) is:

\[
\begin{bmatrix}
\mathbf{f} \\
\mathbf{f}_*
\end{bmatrix}
\sim \mathcal{N} \left( 
\mathbf{0},
\begin{bmatrix}
K(X, X) + \sigma_n^2 I & K(X, X_*) \\
K(X_*, X) & K(X_*, X_*)
\end{bmatrix}
\right)
\]

The posterior predictive distribution for test points \( X_* \) is:

\[
\mathbf{f}_* \mid X, \mathbf{y}, X_* \sim \mathcal{N}(\boldsymbol{\mu}_*, \boldsymbol{\Sigma}_*)
\]

with:

\[
\boldsymbol{\mu}_* = K(X_*, X)[K(X, X) + \sigma_n^2 I]^{-1} \mathbf{y}
\]

\[
\boldsymbol{\Sigma}_* = K(X_*, X_*) - K(X_*, X)[K(X, X) + \sigma_n^2 I]^{-1} K(X, X_*)
\]

\subsection{Kernel Functions}

The kernel function \( k(x, x') \) encodes prior assumptions about the emulated function. One common choice is the RBF kernel:
\[
k(x, x') = \sigma_f^2 \exp\left(-\frac{(x - x')^2}{2\ell^2}\right)
\]
which describes a function that is smooth with a characteristic length scale \( \ell \).  
Kernels often include hyperparameters such as \( \ell \), which are typically optimized by maximizing the marginal likelihood.

\subsection{Advantages and Limitations}

Gaussian Processes provide uncertainty estimates for predictions while being highly flexible. However, their computational cost scales as \( \mathcal{O}(n^3) \) for \( n \) training points, making them inefficient for large datasets. Their performance depends heavily on kernel choice and hyperparameter tuning, so having some prior knowledge of the behavior of the modeled data is advantageous.

Various sparse GP methods, such as inducing point approximations, have been proposed to reduce computational complexity for large datasets.

\acknowledgments

This work uses \texttt{JAX} \citep{jax2018github} and the scientific python stack \citep{jones01, hunter07, vanDerWalt11, scipy}.
Triangle plots have been produced using \getdist{}~\cite{Lewis:2019xzd}.
This project has received funding from the European Research Council (ERC) under the European Union’s Horizon 2020 research and innovation programme (grant agreement No 101001897).
This work has received funding from the Centre National d’Etudes Spatiales and has made use of the Infinity Cluster hosted by the Institut d’Astrophysique de Paris. RKS thanks the Alexander von Humboldt Foundation for their support.
SG and JL acknowledge support from DFG grand LE 3742/6-1. JL and MM acknowledge support from the DFG grant LE 3742/8-1. Calculations for this research were conducted on the Lichtenberg high performance computer of the TU Darmstadt. Computations were performed with computing resources granted by RWTH Aachen University under project `rwth1661'.
We would like to thank Lasse Ausborm for testing \ole{} and Santiago Casas, Jesús Torrado, Andreas Nygaard, Emil Brinch Holm, Steen Hannestad, Will Handley, Thomas Tram, Marco Bonici, Nils Sch\"oneberg and Silvia Galli for useful discussions and comments. SG is first author as the main author of the \ole{} code, remaining authors are in alphabetical order. The authors gratefully acknowledge the computing time provided to them at the NHR Center NHR4CES at RWTH Aachen University (project number p0021792). This is funded by the Federal Ministry of Education and Research, and the state governments participating on the basis of the resolutions of the GWK for national high performance computing at universities (www.nhr-verein.de/unsere-partner).

\bibliography{references}
\end{document}